\documentclass{aa}  

\usepackage{graphicx}
\usepackage{rotating}
\usepackage{xcolor}

\usepackage{txfonts}
\def\BibTeX{{\rm B\kern-.05em{\sc i\kern-.025em b}\kern-.08em
    T\kern-.1667em\lower.7ex\hbox{E}\kern-.125emX}}
\bibpunct{(}{)}{;}{a}{}{,}  

\begin{document}

\title{Carbon stars in the X-shooter Spectral Library \thanks{
Based on observations collected at the European Southern Observatory, Paranal, Chile, Prog. ID 084.B-0869(A/B), 085.B-0751(A/B), 189.B-0925(A/B/C/D).}$^,$ \thanks{Tables~\ref{table_sample},~\ref{table_ext_pp},~\ref{table_color_val},~\ref{table_index_val} 
and spectra are also available in electronic form at the CDS via anonymous ftp to cdsarc.u-strasbg.fr (130.79.128.5) or via http://cdsweb.u-strasbg.fr/cgi-bin/qcat?J/A+A/}}

\author{A. Gonneau\inst{\ref{inst1},\ref{inst2}}
\and A. Lan\c{c}on\inst{\ref{inst1}}
\and S.~C. Trager\inst{\ref{inst2}}
\and B. Aringer\inst{\ref{inst3},\ref{inst4}}
\and M. Lyubenova\inst{\ref{inst2}}
\and W. Nowotny\inst{\ref{inst3}} 
\and R.~F. Peletier\inst{\ref{inst2}}
\and \\ P. Prugniel\inst{\ref{inst5}}
\and Y.-P. Chen\inst{\ref{inst6}}
\and M. Dries\inst{\ref{inst2}}
\and O.~S. Choudhury\inst{\ref{inst7}}
\and J. Falc{\'o}n-Barroso\inst{\ref{inst8},\ref{inst9}}
\and M. Koleva\inst{\ref{inst10}}
\and \\ S. Meneses-Goytia\inst{\ref{inst2}}
\and P. S{\'a}nchez-Bl{\'a}zquez\inst{\ref{inst11}}
\and A. Vazdekis\inst{\ref{inst8},\ref{inst9}}
}

\institute{Observatoire Astronomique de Strasbourg, Universit\'e de
  Strasbourg, CNRS, UMR 7550, 11 rue de l'Universit\'e, \\F-67000 Strasbourg, France\label{inst1}
\and
Kapteyn Astronomical Institute, University of Groningen, Postbus 800, 9700 AV, Groningen, The Netherlands\label{inst2}
\and 
University of Vienna, Department of Astrophysics, T\"urkenschanzstra{\ss}e 17, 1180 Wien, Austria\label{inst3}
\and 
Dipartimento di Fisica e Astronomia Galileo Galilei,
Universit\`a di Padova, Vicolo dell'Osservatorio 3, I-35122 Padova, Italy\label{inst4}
\and
CRAL-Observatoire de Lyon, Universit\'e de Lyon, Lyon I,  CNRS, UMR5574, France \label{inst5}
\and
New York University Abu Dhabi, Abu Dhabi, P.O. Box 129188, Abu Dhabi, United Arab Emirates\label{inst6} 
\and 
Leibniz-Institut f\"ur Astrophysik Potsdam (AIP), An der Sternwarte 16, 14482 Potsdam, Germany \label{inst7}
\and 
Instituto de Astrof\'isica de Canarias, V\'ia L\'actea s/n, La Laguna, Tenerife, Spain\label{inst8}
\and
Departamento de Astrof\'isica, Universidad de La Laguna, E-38205 La Laguna, Tenerife, Spain\label{inst9}
\and
Sterrenkundig Observatorium, Universiteit Gent, Krijgslaan 281, 9000 Gent, Belgium\label{inst10} 
\and
Universidad Aut\'onoma de Madrid, Departamento de F\'isica Te\'orica, E-28049 Cantoblanco, Madrid, Spain\label{inst11}
~\\
\email{anais.gonneau@astro.unistra.fr}
}

\date{Received: 10 April 2015; Accepted: 20 December 2015}

\abstract{We provide a new collection of spectra of 35 carbon stars obtained 
   	with the ESO/VLT X-shooter instrument as
     part of the X-shooter Spectral Library project. The spectra
     extend from 0.3\,$\mu$m to 2.4\,$\mu$m with a
     resolving power above $\sim$ 8000. The sample contains stars
     with a broad range of $(J-K)$ color and pulsation properties
     located in the Milky Way and the Magellanic Clouds. \\
     We show that the distribution of spectral properties of carbon
     stars at a given $(J-K)$ color becomes bimodal (in our sample)
     when $(J-K)$ is larger than about 1.5.  We describe the two
     families of spectra that emerge, characterized by the presence or
     absence of the absorption feature at 1.53\,$\mu$m, generally
     associated with HCN and C$_2$H$_2$.  
	  This feature appears essentially only in large-amplitude va\-ria\-bles,
      though not in all observations. Associated spectral signatures 
      that we interpret as the result of veiling by circumstellar matter,
      indicate that the 1.53\,$\mu$m feature might point to episodes of dust
      production in carbon-rich Miras.}


\maketitle


\section{Introduction}

In the 1860s, Father Angelo Secchi discovered a new type of star -- Type IV
-- known today as carbon stars \citep{Secchi68}. Carbon stars (hereafter C
stars) are on the asymptotic giant branch (AGB) and have spectra
that differ dramatically from those of K- or M-type giants.  C stars
are characterized by spectral bands of carbon compounds, such as CN and
C$_2$ bands, and by the lack of bands from oxides such as TiO and
H$_2$O. The classical distinction between carbon-rich and oxygen-rich
stars is the ratio of carbon to oxygen abundance, C/O. If
$\mathrm{C/O}>1$, oxygen is mostly bound to carbon in the form of
carbon monoxide (CO) because this molecule has a high binding energy.
As a result, little oxygen is left to form other oxides in these
stellar atmospheres, whereas carbon atoms are available to form other
carbon compounds.

Carbon stars are significant contributors to the near-infrared light
of intermediate age stellar populations (1--3\,Gyr)
\citep[e.g.,][]{Ferraro95,Girardi98,Maraston98,Lancon99,Mouhcine_Lancon2,Maraston05,Marigo07}.
The absolute level of this contribution has an impact on mass-to-light ratios
and has important implications for the study of star formation in the
universe.  It is a matter of active research both on the theoretical
side \citep[e.g.,][]{Weiss09,Girardi13,Marigo13} and in the framework of
extragalactic observations
\citep[e.g.,][]{Riffel08,Kriek10,Miner11,Melbourne12,Melbourne13,Boyer13,Zibetti13}.
The qua\-li\-ty of the photometric and spectroscopic predictions made by
po\-pu\-la\-tion synthesis models in this field depends on the existence of
stellar spectral libraries, and their completeness in terms of
evolutionary stages and spectral types.

Carbon stars contribute significantly to the chemical enrichment and
to the infrared light of galaxies, but only small collections of
C-star spectra exist to represent this emission \citep[see][for a
review that includes earlier observations]{Evans10}.  
As a reference for C-star classification, \citet{Barnbaum96} 
published an extensive low-resolution optical spectral atlas (0.4--0.7\,$\mu$m).
It contains 119 spectra. \citet{Joyce98} provided a first impression of 
the near-infrared (NIR) spectra of C stars, again at low spectral resolution 
(48 spectra, with a spectral re\-so\-lu\-tion of $\sim 500$).  
Repeated observations of single long-period variable (LPV)
stars showed significant changes with phase, emphasizing the necessity
of simultaneous observations across the spectrum.  As NIR detectors
improved, \citet{Lancon_Wood} produced a library of 0.5--2.5\,$\mu$m
spectra of luminous cool stars with a resolving power $R=\lambda /
\Delta\lambda \simeq 1100$ in the NIR.  It includes 25 spectra of
seven carbon stars. Simultaneous optical spectra are available for 21
of them, but only at very low resolution ($R \simeq 200$).  More
recently, \citet{Rayner09} have published the IRTF Spectral Library, which
includes 13 stars of spectral type S (C/O = 1) or C. Their spectra have no
optical counterparts, but extend from 0.8\,$\mu$m as far as
5.0\,$\mu$m at a resolving power $R \sim 2000$.

Several population synthesis models have used
the C-star collection of \citet{Lancon_Wood} 
\citep{Lancon99,Mouhcine_Lancon2,Maraston05,Marigo08}.
\citet{Lancon02} suggested using a near-infrared color as a first-order 
classification parameter for the C-star spectra in these
applications, but also noted that this disregards other potentially
important parameters, such as the carbon-to-oxygen (C/O) ratio 
or the pulsation properties. One of the shortcomings of this data 
set is the narrow range of properties \citep{Lyubenova10,Lyubenova12}. 
Another is that it simply contains too few stars to represent the variety 
of C-stars pectra.   

In modeling of luminous cool stellar populations, two important
sources of uncertainties (other than the incompleteness of spectral
libraries) are the fundamental parameters assigned to the observed
stars and the effects of circumstellar dust related to pulsation and
mass loss on the upper asymptotic giant branch. Estimating effective
temperatures, C/O ratios, and gravities requires a comparison with
theoretical spectra. \citet{Loidl01} showed that it is difficult to
obtain a good theoretical representation of both the energy
distribution and the spectral features, even for relatively blue C
stars. \citet{Aringer09} pointed out that static models without
circumstellar dust cannot reproduce any NIR carbon star energy
distribution with $(J-K)>1.6$.  
\citet{Nowotny11,Nowotny13} computed small numbers of spectral energy distributions for
pulsating mo\-dels, at low spectral resolution. 
They reproduce the overall trend from optical carbon stars to dust-enshrouded sources, for
which the whole spectrum is dominated by the emission from dust shells.
But whether or not they reproduce the relationship between color and
the depth of spectral features remains an open question.  It is important to
find out how dust shells may affect the optical and near-infrared
spectra of C stars, especially for objects in the range $ 1.4 \la (J-K) \la 2$
where the NIR luminosities of these AGB stars are large.

In this paper, we present spectra of 35 medium-resolution carbon stars
extending from the near-ultraviolet through the optical into the
near-infrared (0.3--2.5\,$\mu$m). Although this collection is by itself not complete, it
considerably extends the range of data available, and it offers
unprecedented spectral resolution. We expect it to serve both the
purpose of testing theoretical models for C-star spectra (Gonneau et
al., in preparation) and of improving future population synthesis
models.  We describe the input stellar spectral library, our sample
selection and the data reduction in Sections~\ref{section_xsl} and~\ref{section_red}.  
We discuss the spectra using a NIR color as a primary classification criterion in
Section~\ref{section_sample}; in particular we discuss the appearance of a bimodal
behavior of the spectral features and the overall spectral energy
distribution in the redder C-star spectra.  We define a list of
spectroscopic indices in Section~\ref{section_indices} that we use in Section~\ref{section_results} to
quantify the spectral behavior and in Section~\ref{section_comparison} to compare our
spectra with existing libraries of carbon-rich stars.  We discuss our
results in Section~\ref{section_discussion} and present our conclusions in Section~\ref{section_conclusion}.


\section{The XSL carbon star sample}
\label{section_xsl}

With X-shooter \citep{Vernet11}, the European Southern Observatory
(ESO) made available a high-throughput spectrograph allowing the
simultaneous acquisition of spectra from 0.3 to 2.5\,$\mu$m, using two
dichroics to split the beam into three wavelength ranges, referred to as arms:
ultraviolet-blue (UVB), vi\-si\-ble (VIS) and near-infrared (NIR). This simultaneity
is in\-va\-lua\-ble when observing variable stars, and many C stars are LPVs
\citep{Evans10}.

Our team is building a large stellar spectral library under an ESO
Large Programme, the X-shooter Spectral Library \citep[hereafter
XSL,][]{Chen14}. It contains more than 700 stars, observed at moderate
resolving power ($7\,700 \leq R \leq 11\,000$, depending on the arm)
and covering a large range of stellar atmospheric parameters.  The
homogeneous spectroscopic extension into the near-infrared makes XSL
unique among empirical libraries.

In this paper, we focus only on carbon stars. Table~\ref{table_sample} gives a full description of the C-star sample.
Table~\ref{table_ext_pp} summarizes properties of these stars as available in the literature.

\medskip


The sample includes stars from the Milky Way (MW) as well as from the Large and
Small Magellanic Clouds (LMC, SMC).  As C stars on the AGB form a
relatively tight sequence in NIR color-color plots (2MASS,
\citealt{Skrutskie06}; DENIS, \citealt {Epchtein97}; WISE,
\citealt{Wright10}; \citealt{Whitelock06}; \citealt{Nowotny11}), the
primary aim of the selection was to sample an adequate range of
near-infrared colors. This range was restricted to $(J-K)<3$ to avoid
stars with negligible optical flux.  A few C stars with $(J-K)<1$ are
present in the sample, although these stars are considered too hot to
be standard AGB objects; they are instead thought to have become
carbon-rich through other processes, such as mass transfer from a
companion.

While the effects of metallicity on stellar evolution tracks are
large, leading to varying estimates of the fraction of C stars
as a function of metallicity \citep{Mouhcine02,Mouhcine_Lancon,Marigo08,Groenewegen07},
the effects of initial metallicity on the spectrum of a C star of a given
color are relatively small based on static models (\citealt{Loidl01};
Gonneau et al., in prep). Therefore, we initially consider all stars
in the sample as one group, irrespective of the host galaxy.

\medskip

\begin{figure}
	\begin{center}
		\includegraphics[trim=80 50 30 70, clip, width=\hsize]{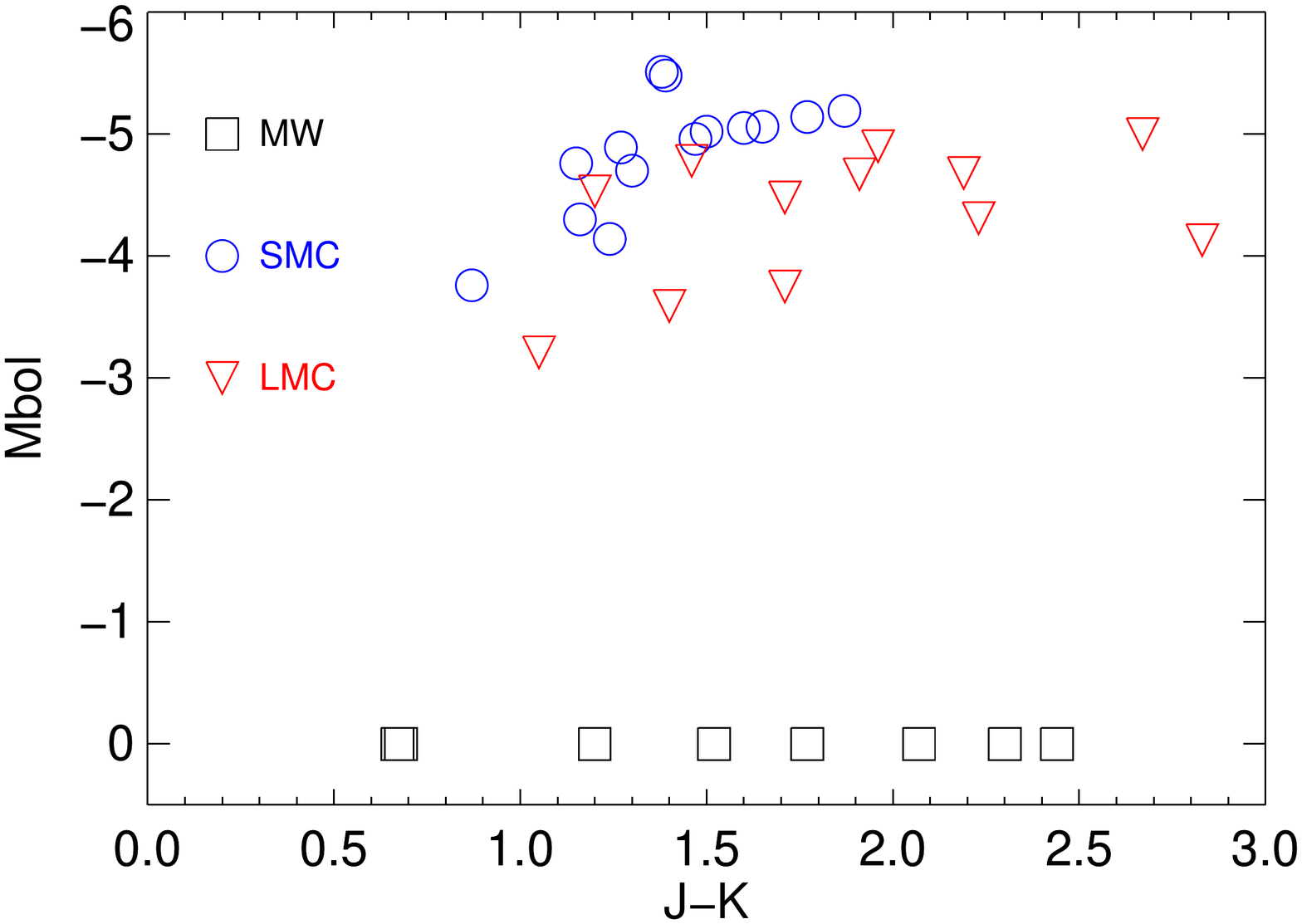}
                \caption{Bolometric magnitudes and literature colors
                  of our sample stars. The LMC stars (red triangles)
                  are taken from \citet{Hughes_Wood}. The SMC stars (blue
                  circles) are derived by \citet{Cioni03}. No reliable
                  distances are known for the MW stars (black
                  squares) of our sample.}
    \label{plot_mbol}
    	\end{center}
\end{figure}
\begin{figure}
	\begin{center}
		\includegraphics[trim=80 50 30 70, clip, width=\hsize]{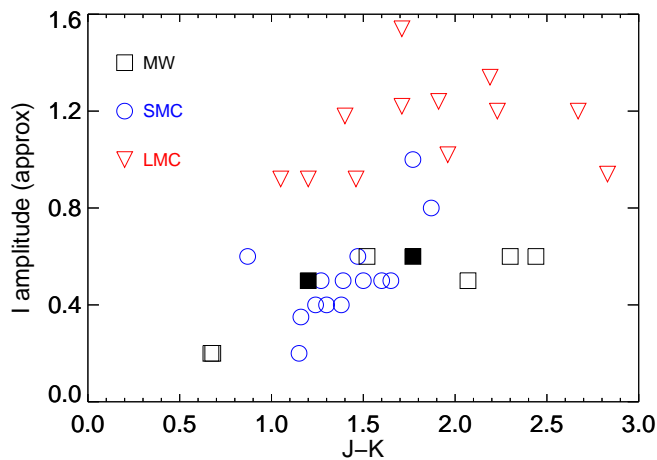}
	 	\caption{$I$-band amplitudes of our sample stars.
                  Symbols are as in Fig.~\ref{plot_mbol}.  The
                  amplitudes are estimated peak-to-peak variations.
                  The values for LMC stars are taken from
                  \cite{Hughes_Wood}.  For the SMC stars, we estimated
                  amplitudes using OGLE light curves (available
                  through the Vizier service at CDS).  For the MW, we
                  estimated amplitudes based on $K$-band amplitudes by
                  \citet{Whitelock06}; a value of 0.5 was assigned
                  when no data were available. We note that for two MW
                  stars (filled squares) large-amplitude 
                  luminosity dips are known to occur
                  occa\-sio\-nally in addition to small-amplitude
                  variations (the R CrB phenomenon).}
    \label{plot_iampl}	
	\end{center}
\end{figure}

A variety of LPV pulsation amplitudes and light curve shapes can be
found in C stars with $1\la(J-K)\la3$. A distinct period and a clear
period-luminosity relation exist for large-amplitude variables
\citep[e.g.,][]{Whitelock06}, but many smaller-amplitude variables
found in this range have extremely irregular light curves
\citep[see][]{Hughes_Wood}.  As shown in Figures~\ref{plot_mbol}
and~\ref{plot_iampl}, there are systematic differences between the
pulsation properties of our subsamples of C stars from the MW, the LMC
and the SMC. On average, the LMC subset has a larger pulsation
amplitude. More specifically, all the LMC stars in our sample display Mira-type
pulsation, while the majority of the other stars in our sample
are semi-regular variables, with comparatively small amplitudes
(Table~\ref{table_ext_pp}). In addition, at a given color, the SMC stars tend to be brighter
than their LMC counterparts (no reliable distances are available for
the MW stars of the sample). These selection biases must be kept in
mind when interpreting the spectra, which is another reason to
  treat the combined samples as one sample.

\begin{table*}
\caption{\label{table_sample}Observational properties of our sample}
\centering
\small
\begin{tabular}{lclccccccc}
\hline\hline
Name & Coordinates & Host & ESO  & ESO	&  MJD	&	Flux 	& $(J-K_s)$  &  Group &  1.53\,$\mu$m \\
	 & (J2000)	   &        & Period  &  OBid  & 		& 	note \tablefootmark{a}		&  [mag]	$^b$		&	$^c$ & 	  feature $^d$\\
\hline
\hline

Cl* NGC 121 T V8 		& 00:26:48.52 $-$71:32:50.5 & SMC & P89 & 723477  & 56090.41 &	N 	&	1.06 & 1 & 	\\
2MASS J00490032-7322238 & 00:49:00.33 $-$73:22:23.8 & SMC & P84 & 389528  & 55110.07 & 	N 	&	1.50 & 2 & \\
2MASS J00493262-7317523 & 00:49:32.62 $-$73:17:52.3 & SMC & P84 & 389526  & 55110.09 &	N 	&	1.44 & 2 & \\
2MASS J00530765-7307477 & 00:53:07.65 $-$73:07:47.8 & SMC & P84 & 389511  & 55116.12 &	N 	&	1.43 & 2 & \\
2MASS J00542265-7301057 & 00:54:22.66 $-$73:01:05.7 & SMC & P84 & 389505  & 55119.07 &	N 	&	1.92 & 3 & \\ 
2MASS J00553091-7310186 & 00:55:30.91 $-$73:10:18.6 & SMC & P84 & 389503  & 55119.09 &	N 	&	2.11 & 3 & \\
2MASS J00563906-7304529 & 00:56:39.06 $-$73:04:53.0 & SMC & P84 & 389499  & 55114.12 &	N 	&	1.37 & 2 & \\
2MASS J00564478-7314347 & 00:56:44.78 $-$73:14:34.7 & SMC & P84 & 389497  & 55119.11 &	N 	&	1.77 & 3 & \\
2MASS J00570070-7307505 & 00:57:00.70 $-$73:07:50.6 & SMC & P84 & 389495  & 55111.07 &	N 	&	1.66 & 3 & \\ 
2MASS J00571214-7307045 & 00:57:12.15 $-$73:07:04.6 & SMC & P84 & 389493  & 55111.08 &	N 	& 	1.53 & 2 & \\
2MASS J00571648-7310527 & 00:57:16.48 $-$73:10:52.8 & SMC & P84 & 389489  & 55111.11 &	N 	& 	1.31	 & 2 & \\
2MASS J01003150-7307237 & 01:00:31.51 $-$73:07:23.7 & SMC & P84 & 389481  & 55111.12 &	N 	& 	1.33	 & 2	& \\
Cl* NGC 419 LE 35 		& 01:08:17.49 $-$72:53:01.3 & SMC & P90 & 804029  & 56213.20 &	V 	& 	2.09	 & 3	& \\
Cl* NGC 419 LE 27 		& 01:08:20.67 $-$72:52:52.0 & SMC & P90 & 804024  & 56213.18 &	V 	& 	1.98	 & 3 & \\

\hline

T Cae 					& 04:47:18.92 $-$36:12:33.5 & MW & P84 & 389388   &	55142.19 &	 N / S	&  1.63	& 3 & \\

\hline

SHV 0500412-684054 		& 05:00:29.71 $-$68:36:37.4 & LMC & P90 & 804254 &  	56213.29 & 		&  1.84 	& 3 &  Y \\
SHV 0502469-692418 		& 05:02:28.86 $-$69:20:09.7 & LMC & P90 & 804257 &  	56213.31 & 		&  1.98 	& 3	& Y \\
SHV 0504353-712622 		& 05:03:55.96 $-$71:22:22.1 & LMC & P84 & 389445 & 	55119.26 & N 	&  2.17	& 3	&  \\ 
SHV 0517337-725738 		& 05:16:33.31 $-$72:54:32.1 & LMC & P90 & 804263 & 	56213.36 &		& 1.13 	& 1 &\\
SHV 0518222-750327 		& 05:16:49.73 $-$75:00:22.7 & LMC & P84 & 389433 &	55142.28 & N 	& 2.52	& 4	& Y\\
SHV 0518161-683543 		& 05:18:02.47 $-$68:32:39.1 & LMC & P90 & 804266 & 	56234.29 & N 	& 	1.16 & 1	& \\
SHV 0520505-705019 		& 05:20:15.02 $-$70:47:26.1 & LMC & P84 & 389428 &	55142.32 & N 	& 	2.37 & 4 & Y \\ 
SHV 0520427-693637 		& 05:20:20.19 $-$69:33:44.8 & LMC & P90 & 804284 &	56240.35 & 		& 	2.11	 & 3 & \\
SHV 0528537-695119 		& 05:28:27.73 $-$69:49:01.8 & LMC & P84 & 389414 & 	55226.19 & V / N  &	3.23	 & 4	& Y \\
SHV 0525478-690944	 	& 05:25:28.21 $-$69:07:13.2 & LMC & P84 & 389421 &	55142.36 & N 	& 	3.02 & 4	& Y \\
SHV 0527072-701238 		& 05:26:37.82 $-$70:10:11.6 & LMC & P90 & 804300 & 	56261.34 & 		& 	2.55 & 4	& Y \\
SHV 0536139-701604 		& 05:35:42.81 $-$70:14:16.3 & LMC & P84 & 389406 &	55226.23 & N 	& 	3.12	 & 4	& Y \\

\hline

{[}ABC89] Pup 42 		& 08:04:57.56 $-$29:51:25.5 & MW & P90 & 804003  &	56292.25 & 	& 2.30	& 4	& \\
IRAS 09484-6242 			& 09:49:49.40 $-$62:56:09.0 & MW & P92 & 998138  &	56617.34 & 	& 2.02	& 3 & \\
{[}W65] c2 				& 11:22:05.06 $-$59:38:45.2 & MW & P90 & 804322  	&	56320.35 & 	& 1.71	& 3 \\
{[}ABC89] Cir 18 		& 13:55:26.20 $-$59:22:19.0 & MW & P89 & 814763  &	56319.37 & 	& 	2.45 	& 4 & \\
 						& 	 				   & MW & P91 & 929000  	&	56383.31 & 	& 	2.52 	& 4 & \\
HE 1428-1950 			& 14:30:59.39 $-$20:03:41.9 & MW & P91 & 929514  	&	56383.33 & 	& 0.71	& 1 	 \\
V CrA\tablefootmark{c}	& 18:47:32.31 $-$38:09:32.3 & MW & P89 & 723829  	&	56144.17 & 		&  -	& - \\
HD 202851 				& 21:18:43.48 $-$01:32:03.3 & MW & P89 & 723822  	&	56144.31 & N	& 0.83 &  1 	& \\
 
\hline
\end{tabular}
\normalsize

\tablefoot{
\tablefoottext{a}{The letter indicates for which X-shooter arm 
\textit{no} absolute flux-calibration was possible: V=visible, N=near-infrared. \\
The S letter indicates that the spectrum is saturated in the $K$-band.}  \\
\tablefoottext{b}{The $(J-K_s)$ colors are derived from the spectra using 
the 2MASS filters \citep{Cohen03} (see Section~\ref{section_color}).}\\
\tablefoottext{c}{The group sharing is discussed in Section~\ref{section_sample}.} \\
\tablefoottext{d}{The Y letter indicates the presence of the 1.53\,$\mu$m absorption band (see Section~\ref{section_sample}).} \\
\tablefoottext{e}{See Appendix~\ref{part_vcra} for more details about V CrA.} 
}
\end{table*}


\section{Data reduction}
\label{section_red}

In the following section, we summarize the applied data reduction
procedures. The carbon star data were acquired over ESO Periods 84,
and 89 to 92 (Table\,\ref{table_sample}). The narrow-slit widths for
UVB, VIS and NIR images were 0.5\arcsec, 0.7\arcsec, 0.6\arcsec, respectively.


\subsection{UVB and VIS arms: extraction and flux-calibration}

The UVB- and VIS-arm carbon star spectra observed in Period 84 are
part of XSL Data Release I \citep[DRI,][]{Chen14} and are used here
unchanged.  The basic data reduction for DRI was performed with
X-shooter pipeline version 1.5.0, up to the crea\-tion of rectified,
wavelength-calibrated two-dimensional (2D) spectra.  The extraction of
one-dimensional (1D) spectra was performed outside of the pipeline
with a procedure inspired by the prescription of \citet{Horne86}.
Observations of both the science targets and spectrophotometric
standard stars through a wide slit (5.00\arcsec) were used to obtain absolute fluxes
(see Table\,\ref{table_sample} for exceptions).

We reduced UVB and VIS spectra from later periods with \mbox{X-shooter}
pipeline version 2.2.0 \citep{Modigliani10}. For the purposes of this
paper, the pipeline was also used for the extraction of 1D spectra and
flux calibration.
The choice of pipeline version does not affect our conclusions.


\subsection{NIR arm: extraction}

All NIR images were reduced with X-shooter pipeline version 2.2.0, up
to the creation of rectified, wavelength-calibrated 2D order spectra.

The extraction of 1D spectra was performed
outside of the pipeline with a procedure of our own. A main driver for
this choice was the need for more control over the rejection of bad
pixels. The standard acquisition procedure for NIR spectra of point
sources is nodding mode, with observations of the target at two
positions (A and B) along the spectrograph slit.  

Instead, we extracted A from (A$-$B) and B from
(B$-$A) and combined them subsequently.  Each extraction implements a
rejection of masked and outlier pixels, as well as a weighting scheme based on a smooth
throughput profile and on the local variance (see Appendix~\ref{part_extraction} for details). 
We note that the spectra in the extreme orders of the NIR arm display some residual curvature and broadening
after pipeline rectification, which our profile accounts for
in a satisfactory fashion. We then merged all the extracted orders to
create a continuous 1D NIR spectrum.

Observations of program stars and of spectrophotometric standard stars
through 5\arcsec-wide slits, required for flux calibration, were
reduced with pipeline sky subtraction switched off because residuals
or negative flux levels were too frequent. The sky was estimated from
both sides of the spectrum at the extraction of 1D spectra.  We did
not implement aperture corrections but used apertures as large as
possible (considering the need to estimate the sky), sometimes at the
expense of the signal-to-noise ratio of these wide-slit spectra.  
A number of the wide-slit observations of carbon stars
lacked significant signal (especially in ESO Period P84; cf.
Table\,\ref{table_sample}), making it impossible to correct the
higher resolution observations of these stars for slit losses.


\subsection{Telluric correction and flux calibration}


X-shooter is a ground-based instrument. Therefore, we must correct our
spectra for extinction by the Earth's atmosphere. Standard flux
calibration procedures account for continuous extinction, but not for
molecular absorption lines (e.g., water vapour, molecular oxygen,
carbon-dioxide, methane). Hereafter, we refer to these as telluric features. 
Telluric absorption features particularly affect the NIR
arm and the reddest part of the VIS arm of X-shooter.

\subsubsection{VIS arm}

The VIS spectra released in DR1 are already flux-calibrated and telluric corrected
\citep{Chen14}. For later periods, the flux calibration was performed within the X-shooter pipeline. The telluric correction was applied afterwards. As for all cool stars in DR1, we selected telluric standard stars (with spectral types B and A) observed close in time and airmass to the carbon stars. We derived the telluric transmission spectra
by removing the intrinsic stellar spectrum, e.g., fitting and removing H lines and normalizing the continuum.
The science spectra were then divided by the transmission spectra.

\begin{figure}
	\begin{center}
		\includegraphics[trim= 80 50 80 70, clip, width=\hsize]{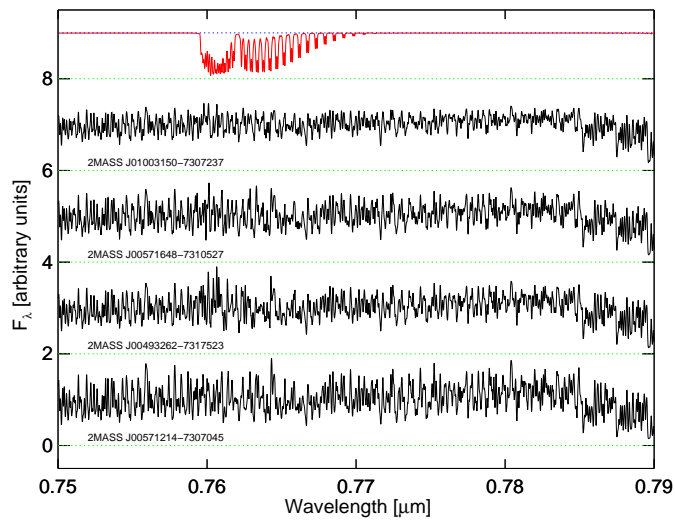}
                \caption{Illustrative spectra in the VIS wavelength
                  range. The red spectrum is a telluric sky model, 
                  smoothed to R $\sim$ 10000.
                  From top to bottom, the stars are 2MASS
                  J01003150-7307237, 2MASS J00571648-7310527, 2MASS
                  J00493262-7317523 and 2MASS J00571214-7307045. 
                  Offsets of 0, 2, 4, 6 and 8 flux units have been applied to the C-star spectra and the telluric spectrum for display.}
    \label{spectra_quality_vis_carbon}
    	\end{center}
\end{figure}

Figure~\ref{spectra_quality_vis_carbon} shows some of the spectra of
the carbon stars over a small part of the visible wavelength range
(0.07\,$\mu$m wide). The red spectrum is a telluric model,
arbitrarily chosen, shown for comparison.

\subsubsection{NIR arm}

The need for specific procedures to account and correct for telluric
absorption is common to many NIR instruments. It is exa\-cer\-ba\-ted in
X-shooter data by an unfortunate feature of the flat-field images. In
the design of the pipeline, spectral features of the flat-field
lamp remain present in the (globally) normalized flat-field images
by which the data are divided and are propagated into the estimated
instrument response curves. One such feature dominates over any other
detector variations, a very strong and sharp bump in the $K$-band
flatfielded spectra, with a much weaker secondary bump in the
$H$-band (see, e.g., Fig.\,9 of \citealt{Moehler14}).  At the altitude
of the ESO Very Large Telescopes, water vapor absorption leaves broad
gaps with no useable data in the NIR spectra, and only very few points
anywhere that are free of any telluric molecular absorption. The
interpolation of estimated response curves through these gaps is a
particularly poorly constrained exercise in the case of X-shooter
pipeline products because only relatively high-order polynomials can
match the bumps due to the flat-field. Therefore, we designed a method
to evaluate the response curve that ex\-pli\-cit\-ly accounts for telluric
absorption. \citet{Moehler14} and \citet{Kausch15} developed in parallel 
other implementations of this idea.

To model the telluric absorption, we chose to use the Cerro Paranal Sky Model, a
set of theoretical telluric transmission spectra provided by J. Vinther and the Innsbruck
team \citep{Noll12,Jones13}. These models are a more complete version
of the spectra that can be found on the web application
SkyCalc\footnote{\url{http://www.eso.org/observing/etc/bin/gen/form?INS.MODE=swspectr+INS.NAME=SKYCALC}}$^,$\footnote{The
  files were computed with version 1.2 based on SM-01 Mod1 Rev.74,
  LBLRTM V12.2, and the line database was aer\_v\_3.2.}. 

The response curve was evaluated as follows. Because the
flat-field bumps are variable in time, we required that the
spectrophotometric standard star used to derive a response curve for a
given program star was reduced with the same flat-field i\-ma\-ges as that
program star. We then fit the flat-fielded spectrum of the flux
standard with the product of the theoretical spectrum of this star, a
telluric transmission model and the unknown response curve. Spectral
regions with telluric features of intermediate depth were used to
select the best-fitting telluric model within the available
collection.  The response curve was represented with a spline
polynomial, with higher concentrations of spline nodes where required
by the flat-fielding bumps.  We corrected the response curve for
continuous atmospheric extinction using the Paranal extinction curve, as available 
in the X-shooter pipeline directory, and taking into account the airmass of the flux
standard.

For the subsequent correction of the narrow-slit spectra of carbon
stars, the search of the ``best'' telluric model is also needed and
more important than above (as we care not only about the shape but also about the lines).  
Therefore, instead of using one telluric model, we allowed for a larger variety of
telluric transmissions by using linear combinations of principal components of the available telluric absorption models, selected within a range of airmasses similar to the airmass of the data.
We performed the $\chi^2$ minimization separately in four wavelength
intervals\footnote{We use the following wavelength regions:
  0.9--1.345\,$\mu$m, 1.46--1.8\,$\mu$m, 1.975--2.1\,$\mu$m and
  2.1--2.5\,$\mu$m.}.  The idea is to better target different
molecules in the telluric spectra.  Then, we divided the science spectrum by the 
telluric transmission and the response curve. We also corrected the science spectrum 
for continuous atmospheric extinction \mbox{using} the airmass at the time of observation.

Whenever possible, the final flux-calibrated spectra were absolutely
flux-calibrated by using wide-slit (5\arcsec) observations (refer to
column ``Flux note'' of Table~\ref{table_sample} for exceptions).

\begin{figure}
	\begin{center}
		\includegraphics[trim= 80 50 80 70, clip, width=\hsize]{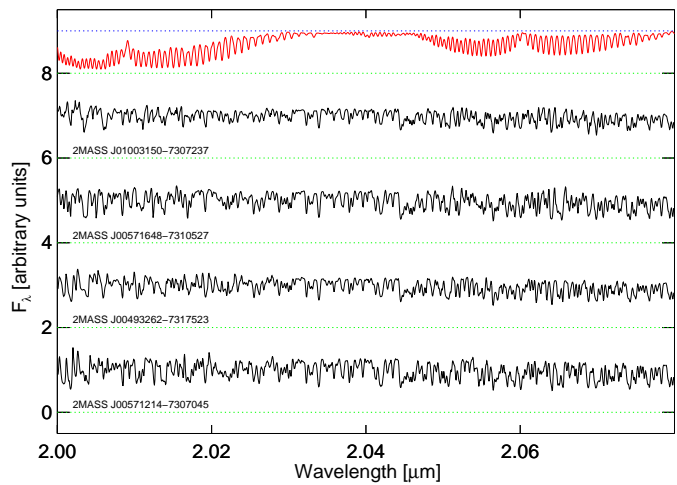}
                \caption{Illustrative spectra in the NIR wavelength
                  range. The red spectrum is a telluric sky model, 
                   smoothed to R $\sim$ 8000. 
                   The stars are the same as in
                  Fig.~\ref{spectra_quality_vis_carbon}.}
    \label{spectra_quality_nir_carbon}
    	\end{center}
\end{figure}

Figure~\ref{spectra_quality_nir_carbon} shows the quality of our
telluric correction process on some of our carbon stars. The red
spectrum shows a telluric model for comparison.

\subsection{Problem with the last order of NIR spectra}

Some of our NIR observations, once flat-fielded, extracted, merged and
flux-calibrated, display a step between the two reddest orders (around
2.27\,$\mu$m, between orders 12 and 11).  This could be related to a
known vignetting problem in order 11, i.e., the last order of the NIR arm that covers 2.28 -- 2.4\,$\mu$m,
to the high background levels in certain exposures, or to other unidentified artifacts. Although we
account for variation in the background emission along the slit in
order 11 and we reduce pairs of one flux standard and one carbon star
with the same flat-field, we cannot eliminate the step completely. 

We correct for the discontinuity, when present, by for\-cing the
average flux level between 2.28 and 2.29\,$\mu$m in order 11 to match
the extrapolation of a linear fit to the spectrum between 2.150 and
2.265\,$\mu$m. This choice is guided by the aspect of theoretical
spectra of C stars \citep{Aringer09} and by published observations with other
instruments \citep{Lancon_Wood, Rayner09}. 
Broad-band colors involving the $K_s$-band change by
(usually much) less than 2\% with this correction.  The estimated
extra uncertainty on measures of the $^{12}$CO bandhead within order
11 remains below a few percent for weak bands, but can reach 10\% for
some of the stars with strong CO bands.


\subsection{Final steps}

We use theoretical models of carbon stars ($R\sim200\,000$), computed
specifically for this paper \citep[based on the atmospheric mo-dels
 of][]{Aringer09}, to shift the wavelength scale of our
observed spectra to the vacuum rest-frame.

Finally, the three arms are merged to produce a complete spectrum
from the near-UVB to the NIR \footnote{The three arms overlap quite
  well: UVB: 0.3--0.59\,$\mu$m; VIS: 0.53--1.02\,$\mu$m; NIR:
  0.99--2.48\,$\mu$m.}.  The resolving power in the UVB, VIS and NIR
ranges of the merged spectra are, res\-pec\-ti\-ve\-ly, $R\sim9100$, $\sim 11
000$ and $\sim 7770$.


\section{The diversity of carbon star spectra}
\label{section_sample}



Our sample of carbon stars presents quite a diversity in spectral
shape and absorption-line characteristics. Figures~\ref{all_carbon_p1}
to~\ref{all_carbon_p6} show our spectra from the UVB to the NIR wavelength
range.  The spectra were normalized to the flux at 1.7\,$\mu$m and shifted for
display purposes. The gray bands in the NIR mask regions where the
telluric absorption is deepest and the signal cannot be recovered. It
is important to note that the spectra were heavily smoothed to a
common resolution (R $\sim$ 2000) in these figures.

\begin{figure*}
	\begin{center}
		\includegraphics[trim= 40 70 60 100, width=\textwidth]{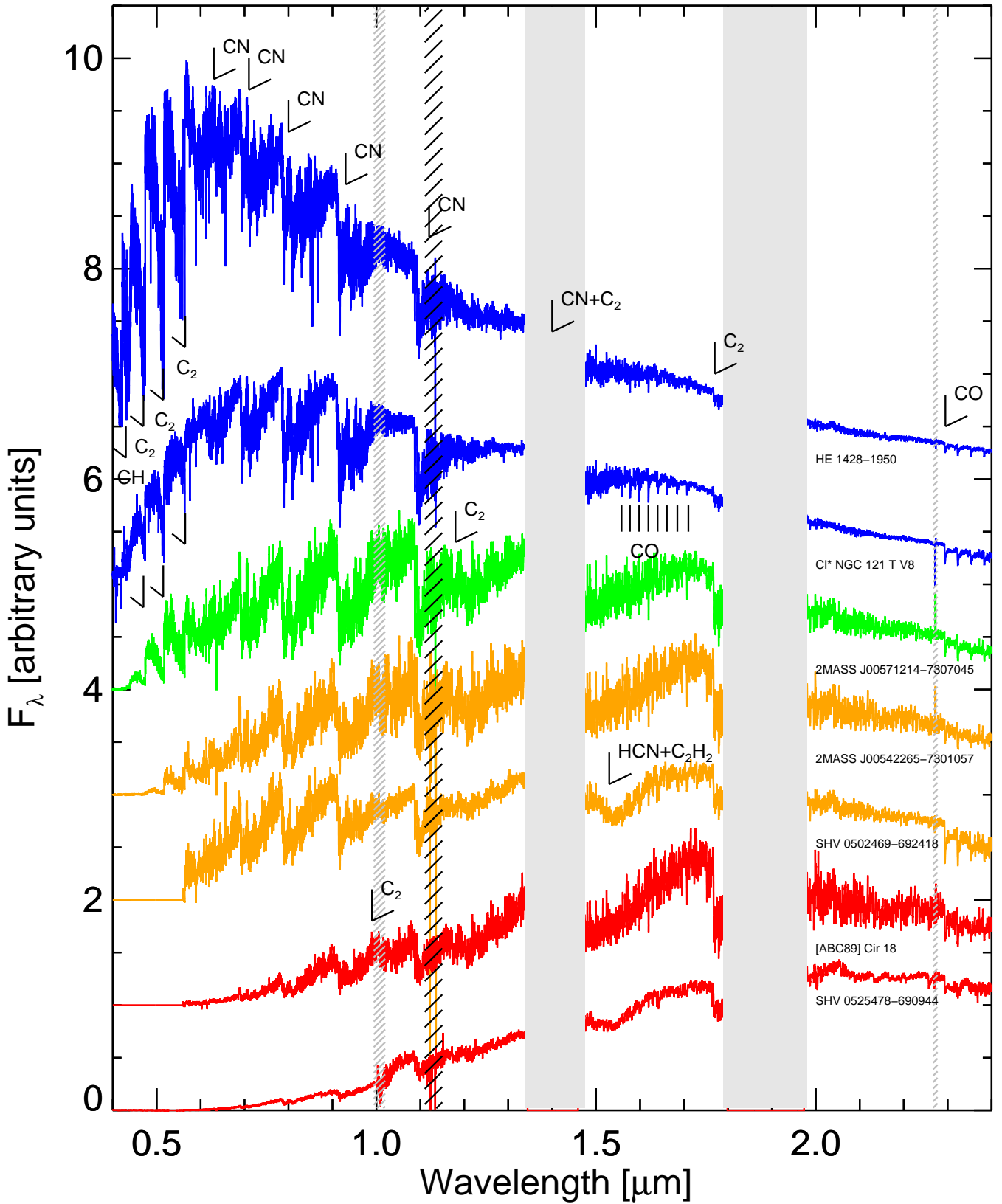}
                \caption{Representative spectra from our sample of carbon stars. 
                	The gray bands mask the regions where telluric absorption is strongest. 
                	The areas hatched in black are those that could not be corrected for telluric absorption in a satisfactory way. 
					The areas hatched in gray are the merging regions between the VIS and NIR spectra and between the last two orders of the NIR spectrum.        
					In some spectra, data are missing at 0.635\,$\mu$m.       	
                	The spectra have been smoothed for display purposes to R $\sim$ 2000. 
                	The colors represent the different groups of carbon stars. 
                	Group 1 stars are shown in blue (from top to bottom: HE
                  1428-1950 and Cl* NGC 121 T V8), 
                  Group 2 in red (2MASS J00571214-7307045),               
                  Group 3 in orange (from top to bottom: 2MASS J00542265-7301057
                  and SHV 0502469-692418) and Group 4 in red (from top to bottom: [ABC89] Cir 18 and SHV 0525478-690944). }
	     \label{spectra_carbon}
    	\end{center}
\end{figure*}

We group our carbon stars by $(J-K)$ values when describing them in
the remainder of this Section. Figure~\ref{spectra_carbon} summarizes
the spectral variety of the sample, showing representative examples of
each group. The colors are used to better identify the different
groups.

CO, CN and C$_2$ produce the vast majority of features 
in spectra of carbon stars in this wavelength range.
Some bands have sharp bandheads, but a forest of lines
from various transitions are spread across the whole spectrum.
The C$_2$ Swan bands \citep{Swan57} are dominant between 0.4 and 0.6\,$\mu$m.
Longward of 0.6\,$\mu$m, the most prominent bandheads are
due to CN. Note that the NIR CN bands, in particular the 1.1\,$\mu$m
bandhead, are also seen in M giants and supergiants
\citep{Lancon_Hauschildt,Davies13}.  The C$_2$ band at 1.77\,$\mu$m is
one of the unam\-bi\-guous characteristics of C\,stars in the NIR.  The CO
bands in the $H$ and $K$ windows are also present with varying strengths in
all the spectra.

First we describe spectral characteristics of each group seen 
from visual inspection. We perform a more quantitative a\-na\-ly\-sis in subsequent sections.


\subsection{Group 1 -- The bluest stars: $(J-K)< 1.2$ [5 stars]}
\label{blue_stars}

Figure~\ref{all_carbon_p1} shows the five warmest C stars in our
sample. The top two spectra of Fig.~\ref{spectra_carbon}, displayed in
blue, are representative of the two types of behaviors found in this
group.

The top two stars of Fig.~\ref{all_carbon_p1}, HD 202851 and HE
1428-1950, have spectra similar to those of early K type giants (CN
band at 0.431\,$\mu$m and G band of CH of similar strength, H$\beta$
line in absorption). But they clearly display the C$_2$ bands
characte-ristic of C stars, in particular the Swan bands around
0.47\,$\mu$m and 0.515\,$\mu$m.

The three bottom spectra of Fig.~\ref{all_carbon_p1} have spectral
energy distributions (SED) that peak at longer wavelengths, but have weaker
molecular features in the optical range. The CN band at 0.431\,$\mu$m
and the G band of CH are undetectable in two of the three stars.
On the other hand, the red system of
CN is slightly stronger, and the NIR C$_2$ bandhead (1.77\,$\mu$m) and
the CO bands are significantly stronger. Two of these three spectra
display hydrogen emission lines, a phase-dependent signature of
pulsation.


\subsection{Group 2 -- Classical C stars : $1.2<(J-K)<1.6$ [7 stars]}

Figure~\ref{all_carbon_p2} shows classical C stars: seven stars belong
to this group. The third spectrum in Fig.~\ref{spectra_carbon}
(2MASS J00571214-7307045), displayed in green, is representative of this group.  
All these C stars have $1.2<(J-K)<1.6$. They happen to be located in the SMC, but
we note that many of the Galactic C stars of \cite{Lancon_Wood}
would fall in this category.

At optical wavelengths, the Swan bands are the first features to
appear when $\mathrm{C/O} > 1$. Compared to the first group of
spectra, the spectra collected here have significantly stronger
absorption bands of CN and C$_2$ in the NIR. C$_2$ absorption modifies
the spectrum across the $J$ band and creates a strong bandhead at
1.77\,$\mu$m.  A forest of lines of both CN and C$_2$ is responsible for
the rugged appearance of the spectrum, which should not be mistaken as an
indication of noise. CO bands in the $H$ window appear weak, as a combined consequence
of the C/O ratio and of overlap with many other features.

The SED and the H band (CO, C$_2$, forest of CN and C$_2$ lines) of
the top spectrum of Fig.~\ref{all_carbon_p2} (2MASS J01003150-7307237) seem to indicate that this star
is slightly warmer, or has a lower C/O ratio, than the others. The
opposite holds for the last spectrum in that figure.


\subsection{Group 3 -- Redder stars: $1.6<(J-K)<2.2$ [13 stars]}

Figures~\ref{all_carbon_p3} and~\ref{all_carbon_p4} show redder stars.
Thirteen stars compose this group. Group 3 is less homogeneous than
Group 2: while some spectra simply seem to extend the sequence of
Group 2 to redder SEDs with stronger features, others deviate from
this behavior. This leads us to define two subgroups. A
representative of each subgroup is included in
Fig.~\ref{spectra_carbon} (see the fourth and fifth spectra, displayed
in orange).

The stars that simply extend the behavior of group 2 have strong
C$_2$ bands in the $J$- and $H$-bands and weak CO bands.  In two of
these stars (T Cae and [W65] c2), the CO bands are re-latively stronger,
suggesting a C/O ratio closer to 1 \citep[cf. the S/C star BH\,Cru
in][]{Lancon_Wood}. We warn however that the interpretation of ratios
of CO to other band strengths in terms of abundance ratios is a
non-trivial exercise, as the band-strength ratios may depend on phase
\citetext{\citealp[see the multiple spectra of R\,Lep
  by][]{Lancon_Wood}, \citealp[or those of V\,Cyg from][]{Joyce98}}.

Two stars stand out: SHV 0500412-684054 and
SHV 0502469-692418.  They are characterized by an absorption band
around 1.53\,$\mu$m (see Section~\ref{part_153_feature}), weaker C$_2$ absorption, some of the strongest 
2.3\,$\mu$m CO absorption of this group, and a significantly smoother general appearance
than other spectra. This latter property has, to
our knowledge, never been emphasized before.  In hindsight, it is
also noticeable in previously published spectra that display the
1.53\,$\mu$m feature \citep{Lancon_Wood,Groenewegen09,Rayner09}.

Inspection of the spectra with the 1.53\,$\mu$m feature also suggests
that their emission in the red part of the optical spectra is
relatively strong, considering their red NIR spectra. The spectra
however drop rapidly to a negligible flux in the blue.


\subsection{Group 4 -- The reddest stars: $(J-K)>2.2$ [9 stars]}

The last two figures, Figures~\ref{all_carbon_p5}
and~\ref{all_carbon_p6}, show our reddest stars. The dichotomy seen in
Group 3 is obviously present here as well. The two bottom
spectra from Fig.~\ref{spectra_carbon}, displayed in red, are
representative of that group, composed of nine stars.  The general
trend in this group is that the energy peak shifts from the optical to
the near-infrared, which leads to a ``plateau'' in the $K$ band of the
reddest objects.

It is interesting to note that at higher $(J-K)$, more stars display
the 1.53\,$\mu$m absorption feature. The fact that the feature appears
only in very red stars (but not in all very red stars) is consistent
with previous observations \citep{Joyce98,Groenewegen09}.  The spectra
that display 1.53\,$\mu$m absorption share the properties already
mentioned for their counterparts in Group 3. They clearly have a
smoother appearance than the others. Unfortunately, the S/N ratio is
poor for some of these objects beyond 2.25\,$\mu$m.  Where it is good,
inspection by eye indicates that the CO bands in these smoother spectra
have strengths similar to those in spectra with a forest of CN and
C$_2$ lines.  In three cases, SHV\,0520505-705019,
SHV\,0527072-701238, SHV\,0528537-695119, there seems to be an excess
of flux in the red part of the optical spectrum, compared to other
stars.


\subsection{The 1.53\,$\mu$m feature}
\label{part_153_feature}

The 1.53\,$\mu$m absorption feature was first noticed in cool
carbon stars of the Milky Way \citep{Goebel81,Joyce98}.
These authors associated it with large-amplitude variability.
The underlying molecules are most likely a combination of HCN and
C$_2$H$_2$ \citep{Loidl04}, and the 1.53\,$\mu$m band is thought to be
an overtone of the strong absorption band sometimes seen around
3\,$\mu$m (e.g., in the spectrum of R\,Lep in the IRTF library). 
However, this correlation remains to be formally established, as \citet{Joyce98}
stated that this correlation is poor.

In our sample, all the spectra that display the 1.53\,$\mu$m feature
belong to large-amplitude variables. The fact that in our sample
they all happen to
be in the LMC should be seen as a selection effect, since stars with
this feature have previously been found both in the Milky Way and 
in dwarf galaxies more metal-poor than the LMC \citep[e.g.,
the Sculptor dwarf, ][]{Groenewegen09}.


\section{Spectroscopic indices for carbon stars}
\label{section_indices}

We derive spectroscopic indices from our flux-calibrated spectra to compare 
our sample with previous studies. 
Unless otherwise stated, we use the following formula to measure the strength 
of any absorption band $X$:
\begin{equation}
I(X) = -2.5 \log_{10}[F_b(X) / F_c(X)],
\label{eq_index}
\end{equation}
where $F_b(X)$ and $F_c(X)$ are the mean energy densities received in the
wavelength bin in the absorption band region and the
(pseudo-)``continuum''\footnote{
At the resolution of our C-star spectra, the true continuum is inaccessible anywhere. 
What we call pseudo-``continuum'' in this section is simply a reference flux level measured outside 
the molecular band of interest for a particular spectrophotometric index, following earlier usage by, e.g., \citet{Worthey94}.} of index $X$.

We note here that these ``one-sided'' indices depend on the quality of
the flux calibration over moderate wavelength spans, in contrast to
the classic ``two-sided'' Lick/IDS-type indices such as those defined
by, e.g., \citet{Burstein84}, which are robust to small-scale
flux-calibration issues. 
We evaluate error bars on indices by measuring them on spectra
reduced with several estimates of the instrumental response curve, 
and by taking the standard deviation of these measurements.

Table~\ref{table_filters} summarizes the properties of the bandpasses
used to define our spectroscopic indices, as illustrated in Figures~\ref{index_CO12}--\ref{index_C2U}.


\begin{table*}
\caption{\label{table_filters} Properties of our spectroscopic indices}
\centering
\begin{tabular}{llcclcc}
\hline
Index & Bandpass feature	& $\lambda_{min}$ ($\mu$m) & $\lambda_{max}$ ($\mu$m) 	&	Bandpass  ``continuum''	& $\lambda_{min}$ ($\mu$m) & $\lambda_{max}$ ($\mu$m)  \\
\hline

C2U 	& C2U\_band		& 0.5087 & 0.5167					& C2U\_cont 	& 0.5187 & 0.5267 \\

CN		& W110 				& 1.0970 & 1.1030						& W108 		& 1.0770 & 1.0830  \\

DIP153 	& DIP153\_band 			& 1.5000 & 1.6000 					& DIP153\_cont 	& 1.4800 & 1.5000 		\\

COH$_{52}$ 	& COH52\_band		& 1.5974 & 1.6026			 		& COH52\_cont	& 1.5914 & 1.5966		\\
COH$_{63}$ 	& COH63\_band 		& 1.6174 & 1.6226 					& COH63\_cont	& 1.6114 & 1.6166		\\

C2		& C2\_band			& 1.7680 & 1.7820				 		& C2\_cont 	& 1.7520 & 1.7620 \\

CO12 	& KH86CO1 			& 2.2931 &  2.2983					& KH86c1 	& 2.2873 & 2.2925 \\

CO13 	& KH86CO3 			& 	2.3436 & 2.3488 				& KH86CO2 	& 2.3358 & 2.3410 \\

\hline

\end{tabular}
\end{table*}


\subsection{Broad-band colors}
\label{section_color}

For each of our spectra, we derived synthetic magnitudes using the
\citet{Bessel90} filters $R$ and $I$ and the 2MASS filters
\citep{Cohen03} $J$, $H$ and $K_s$. We use these magnitudes to define
the colors $(R-H)$, $(R-I)$, $(I-H)$, $(I-K)$, $(J-H)$, $(H-K)$ and
$(J-K_s)$.

\begin{figure}
	\begin{center}
		\includegraphics[trim=65 30 75 70, clip, width=\hsize]{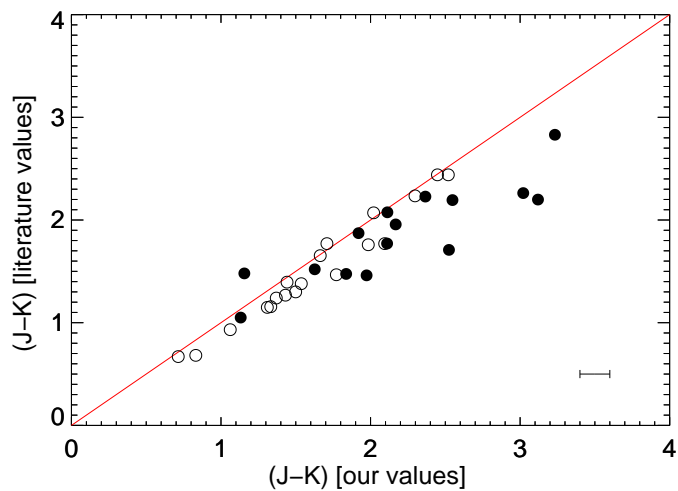}
                \caption{Comparison of our $(J-K_s)$ values with
                  values found in the li\-te\-ra\-tu\-re \citep[2MASS
                  values,][]{Cutri03}. The red line indicates the
                  one-to-one relation. The black points represent the
                  stars with large-amplitude ($I$ amplitude $\geq 0.8$). 
                  The bar plotted in the bottom-right corner shows the $\pm1\sigma$
                  root-mean-square deviation
                  of our photometry with respect to the literature (large-amplitude variables 						  excluded). This is an upper limit of the uncertainties 
                  in the flux calibration and any possible residual variability.}
    \label{plot_comp_litt}
    	\end{center}
\end{figure}

Figure~\ref{plot_comp_litt} compares our $(J-K_s)$ colors with those
found in the literature. 
When we exclude large-amplitude variables, we 
find a good agreement\footnote{The offset might be due to zero-point differences 
between our synthetic photometry and the listed 2MASS magnitudes. Indeed, using the reference 
fluxes for zero-magnitude stars in \citet{Cohen92} and a recent template Vega spectrum 
from the Hubble Space Telescope calibration documentation (\url{ftp://ftp.stsci.edu/cdbs/current_calspec/}), 
our synthetic photometry gives $(J-K_s)=0$ for Vega, 
while the 2MASS point source catalog lists $(J-K_s)=-0.31$ for that star. }. 
The dispersion increases with redder colors, as already noted by \citet{Whitelock09}.
The large scatter for the large-amplitude pulsators is not sur\-pri\-sing, 
since we are com\-pa\-ring our instantaneous color with 2MASS observations
about ten years old, and many light curves display long term trends
in addition to periodicity.

\subsection{$^{12}$CO indices}

\begin{figure}
	\begin{center}
		\includegraphics[trim= 80 30 75 70, clip, width=\hsize]{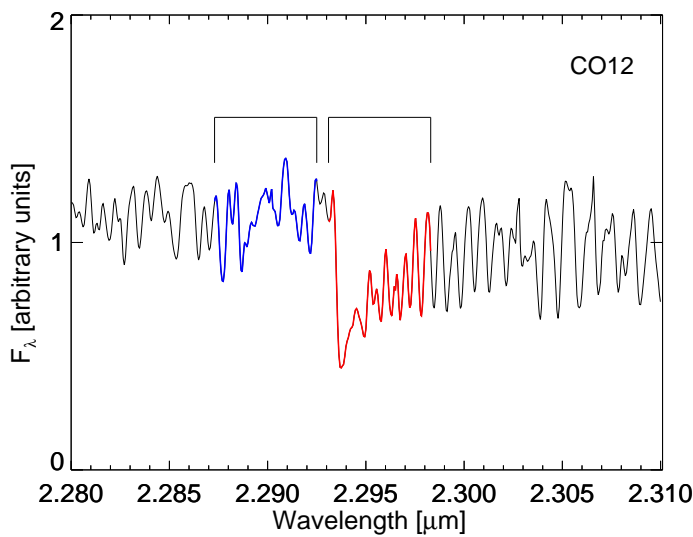}
                \caption{Zoom into the $^{12}$CO(2,0) feature near
                  2.3\,$\mu$m. The red line indicates the region used
                  to calculate the KH86CO1 bandpass, while the blue
                  line corresponds to the KH86c1 bandpass measuring
                  the continuum. This star is 2MASS
                  J00571648-7310527.}
    \label{index_CO12}
    	\end{center}
\end{figure}

We first look at the CO bands located in the $K$ band. To measure the
$^{12}$CO(2,0) bandhead around 2.29\,$\mu$m, we use the definition
given by \citet{KH86}: the absorption bandpass is centered on 2.295\,$\mu$m 
(\textit{KH86c1}) and the continuum bandpass is centered on 2.2899\,$\mu$m (\textit{KH86CO1}).
We call this index \textit{CO12}.
Figure~\ref{index_CO12} shows these passbands on one of our spectra.

\medskip

\begin{figure}
	\begin{center}
		\includegraphics[trim=55 20 60 50, clip, width=\hsize]{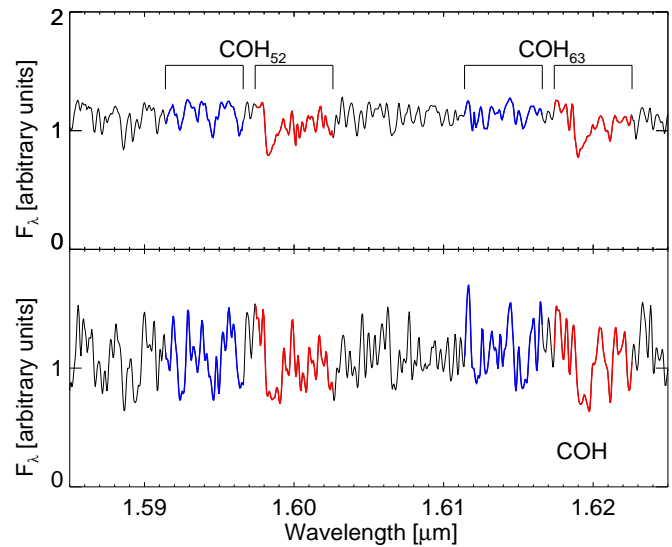}
                \caption{Zoom into the $^{12}$CO(6,3) line near
                  1.62\,$\mu$m. The red lines indicate the features
                  regions, while the blue measure the continuum.
                  The upper panel shows the spectrum of a carbon star, 
                  Cl* NGC 121 T V8, in which the
                  $^{12}$CO(5,2) and $^{12}$CO(6,3) bands are visible.
                  The lower panel corres\-ponds to another carbon star,
                  2MASS J00571214-7307045, in which the CN and C$_2$
                  lines are more prominent and overlap with the CO
                  lines. The spectra have been normalized at
                  1.62\,$\mu$m for display purposes.}
    \label{index_COH}
    	\end{center}
\end{figure}

Other CO bands can also be found in the C-star spectra.
\citet{Origlia93} studied the CO band in the $H$-band near
1.62\,$\mu$m, corresponding to $\Delta \upsilon = 3$ ro-vibrational
bands. In our spectra, this $^{12}$CO(6,3) line does not always
appear. Figure~\ref{index_COH} shows two examples taken from our
sample where the CO(6,3) lines are seen (upper panel) or are hidden under
a combination of CN and C$_2$ lines (lower panel). 

We define a new
index \textit{COH}, based on two other indices measuring the $^{12}$CO(5,2) at 1.60\,$\mu$m ($COH_{52}$)
and the $^{12}$CO(6,3) line at 1.62\,$\mu$m ($COH_{63}$):
\begin{equation}
COH = (COH_{52}+COH_{63})/2.
\end{equation}

\subsection{$^{13}$CO index}

\begin{figure}
	\begin{center}
		\includegraphics[trim=80 30 75 70, clip, width=\hsize]{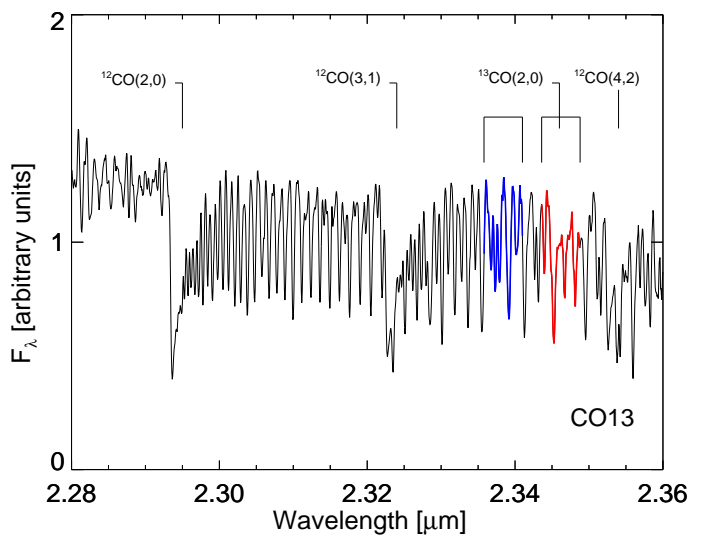}
                \caption{Zoom into the $^{13}$CO(2,0) line near
                  2.35\,$\mu$m. The red line indicates the region used
                  to calculate the KH86CO3 bandpass, while the blue
                  line corresponds to the KH86CO2 bandpass measuring
                  the continuum. This star is SHV 0500412-684054.}
    \label{index_CO13}
    	\end{center}
\end{figure}

The strongest $^{13}$CO features in the spectra are the $\Delta \upsilon = 2$ ro-vibrational bands located around 2.53\,$\mu$m. To measure the $^{13}$CO(2,0) bandhead, we use the definition of the
bandpass centered on 2.3462\,$\mu$m given by \citet{KH86}
(\textit{KH86CO3}). We do not use the same definition for the
con\-ti\-nuum as their bandpass is centered on 2.2899\,$\mu$m and thus
too far away from the absorption bandpass and more likely to be
sensitive to slope effects. Therefore, we define a new bandpass for
the continuum and create the index \textit{CO13}.
Figure~\ref{index_CO13} shows the definition of the bands used to
define this index on one of our spectra.

\subsection{CN index}

\begin{figure}
	\begin{center}
		\includegraphics[trim=80 30 75 70, clip, width=\hsize]{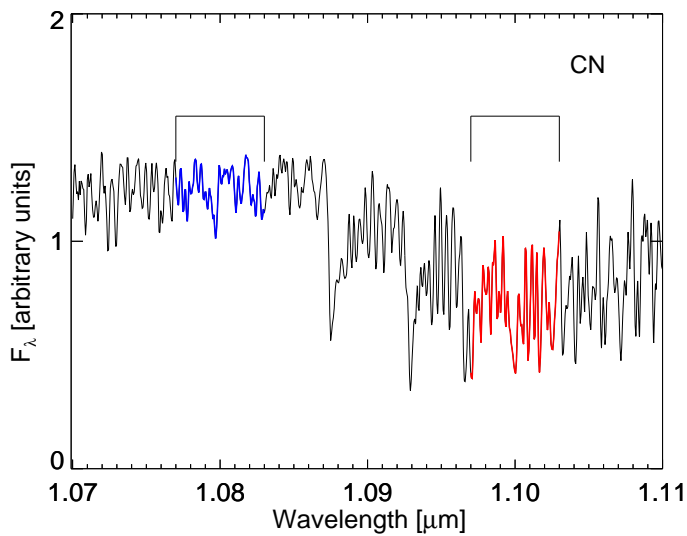}
                \caption{Zoom into the CN line near 1.1\,$\mu$m. The
                  red line indicates the region used to calculate the
                  W110 bandpass, while the blue line corresponds to
                  the W108 bandpass measuring the continuum. This
                  star is SHV 0502469-692418.}
    \label{index_CN}
    	\end{center}
\end{figure}

CN is also seen in carbon-rich spectra. The red system of CN 
appears beyond 0.5\,$\mu$m and the bands grow towards
longer wavelengths. To estimate the CN in the NIR part of the
spectrum, we use rectangular filters adapted from the 8-color system of \citet{Wing71}.
We use an index \textit{CN}, based on \citeauthor{Wing71}'s W110 and W108 passbands, that measures the CN
feature at 1.1\,$\mu$m. Figure~\ref{index_CN} shows these passbands.

\subsection{Measure of the 1.53 $\mu$m feature}

\begin{figure}
	\begin{center}
		\includegraphics[trim=45 20 60 50, clip, width=\hsize]{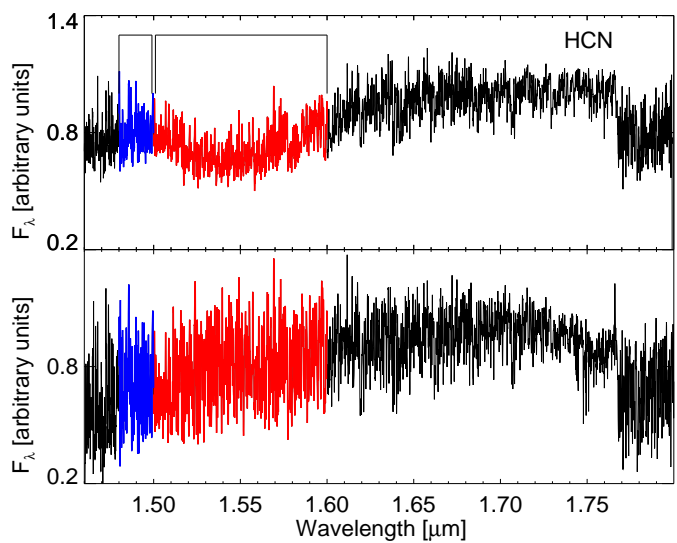}
                \caption{Zoom into the HCN+C$_2$H$_2$ lines around
                  1.53\,$\mu$m. The upper panel shows an example of a
                  carbon star, SHV 0502469-692418, in which the
                  HCN+C$_2$H$_2$ feature is visible. The lower panel
                  corresponds to another carbon star, T Cae, in which
                  the feature is missing. The red line indicates the
                  region used to calculate the absorption bandpass,
                  while the blue line corresponds to the continuum
                  bandpass.}
    \label{index_HCN}
    	\end{center}
\end{figure}

Some of our stars exhibit the 1.53\,$\mu$m feature. We interpret this
feature to be caused by HCN+C$_2$H$_2$.  We define an index \textit{DIP153} to
measure its depth. Figure~\ref{index_HCN} shows two examples taken
from our sample where this feature is seen (upper panel) and absent
(lower panel).

\subsection{C$_2$ indices}

C$_2$ bands appear at many wavelengths in the spectra of carbon stars. The bands
between 0.4 and 0.7\,$\mu$m correspond to the Swan system
\citep{Swan57}. The C$_2$ bands at 0.77, 0.88, 1.02 and 1.20\,$\mu$m
are part of the Phillips system \citep{Phillips48}. 

\begin{figure}
	\begin{center}
		\includegraphics[trim=80 30 75 70, clip, width=\hsize]{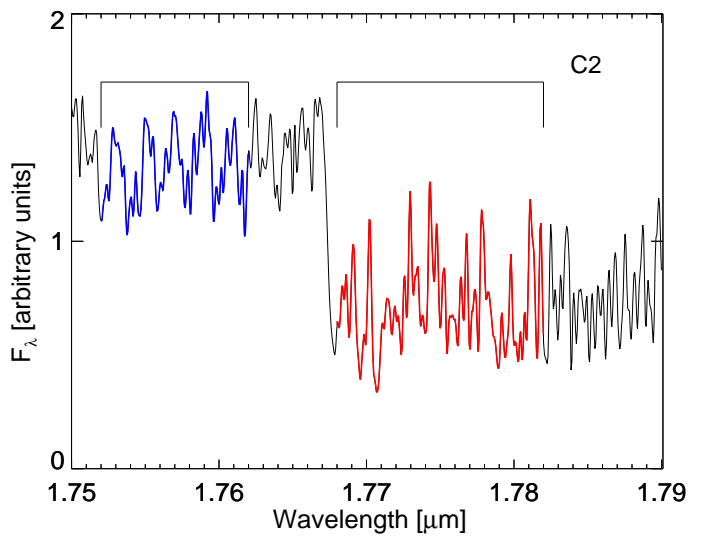}
                \caption{Zoom into the C$_2$ line around 1.77\,$\mu$m.
                  The red line indicates the region used to calculate
                  the \textit{C2\_band} bandpass, while the blue line
                  corresponds to the bandpass measuring the
                  continuum. This star is 2MASS
                  J00563906-7304529.}
    \label{index_C2}
    	\end{center}
\end{figure}

The Ballik-Ramsay fundamental band is the C$_2$ band at
1.77\,$\mu$m \citep{Ballik_Ramsay}. To measure it, we use
the same definition as \citet{Alvarez00}. The bandpasses used to
define the \textit{C2} index are shown in Figure~\ref{index_C2}.

\begin{figure}
	\begin{center}
		\includegraphics[trim=30 20 60 50, clip, width=\hsize]{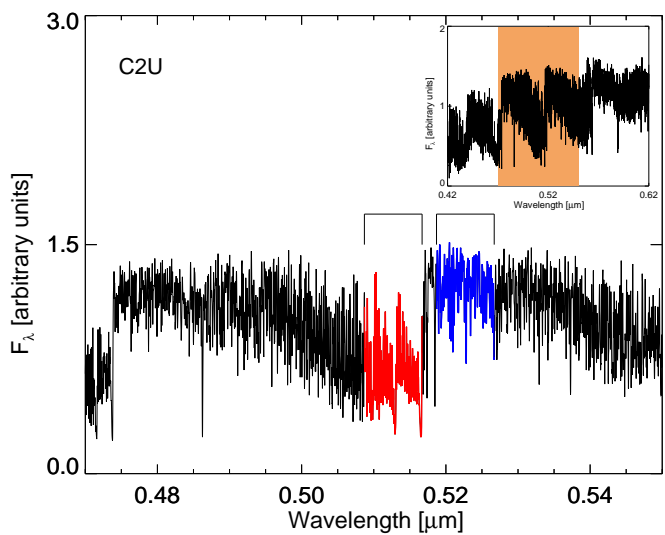}
                \caption{The C$_2$ Swan system (top panel) and a zoom
                  into the C$_2$ line around 0.5165\,$\mu$m. The red
                  line indicates the region used to calculate the
                  \textit{C2U\_band} bandpass, while the blue line
                  corresponds to the \textit{C2U\_cont} bandpass
                  measuring the continuum. This star is HE
                  1428-1950.}
    \label{index_C2U}
    	\end{center}
\end{figure}

The upper-right panel of Figure~\ref{index_C2U} shows the C$_2$ Swan
system for one of our spectra. Due to low signal-to-noise for the XSL
carbon stars at short wavelengths, the bands near 0.47\,$\mu$m or bluer are barely seen.
The band near 0.56\,$\mu$m is problematic because of instrumental
issues in that range \citep{Chen14}, and the bands above 0.6\,$\mu$m
are too heavily contaminated by CN. This leaves the band near
0.5165\,$\mu$m as our best choice to define an index, even though this
index cannot be defined for all stars of our sample due to an absence
of signal in the UVB/VIS parts of some of our spectra. We call this
index \textit{C2U}.

\subsection{Measure of the high-frequency structure}

We also estimate the high-frequency structure in the $H$ and $K$
bands, inspired by the smooth appearance of the NIR spectra of C stars
with the 1.53\,$\mu$m feature.  For both windows, we first fit a straight line
on the wavelength range under study: 1.66 -- 1.7\,$\mu$m for the $H$
band and 2.18 -- 2.23\,$\mu$m for the $K$ band. We then divide our
spectrum by this fit, thus normalizing the continuum. Next, for any
window $X$, we derive the rms from the following formula:
\begin{equation}
rms(X) = \sqrt{\frac{\sum_{i}^{N}(X_i-1)^2}{N}}.
\label{eq_rms}
\end{equation}


\section{Results}
\label{section_results}


\subsection{Color-color plots}

\begin{figure*}
	\begin{center}
		\includegraphics[trim=40 40 0 430]{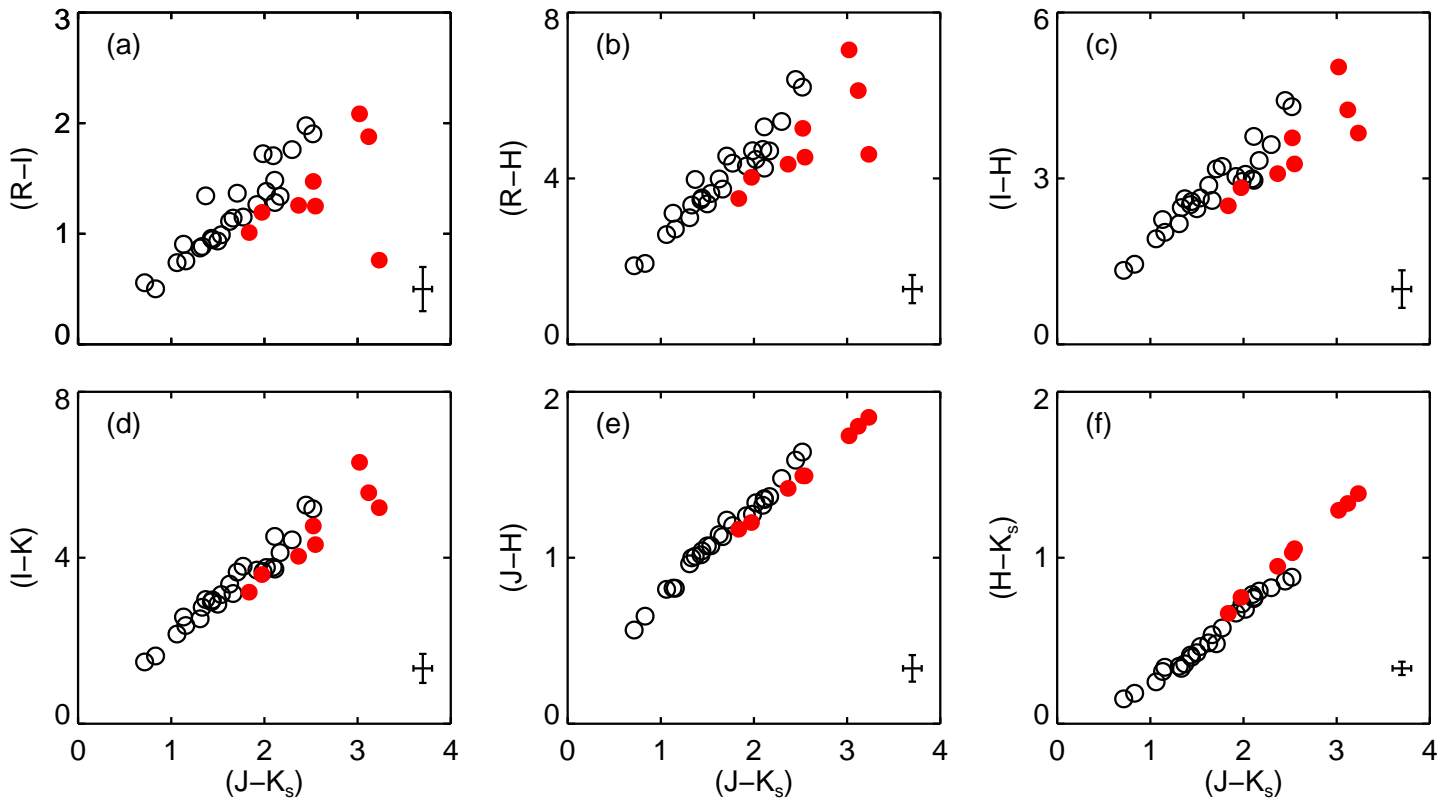}
                \caption{Some color-color plots derived from our
                  sample of carbon stars. The red circles stand for
                  carbon stars showing the 1.53\,$\mu$m feature.              
				  For the $R$ filter, we adopt some conservative values. 
				  	The bars plotted in the bottom-right corners are determined as in Figure~\ref{plot_comp_litt}.}
	     \label{plot_color_color}
    	\end{center}
\end{figure*}

Figure~\ref{plot_color_color} shows the locus of the observed carbon stars
in color-color diagrams, based on the measurements made in Section~\ref{section_color}. 
On the whole, different colors are well correlated with each other. 
The dispersion along the trend is smaller at bluer co\-lors.

The red circles stand for carbon stars showing the 1.53\,$\mu$m feature
in their spectra.
A trend characterizes these stars in the color-color
diagrams: at a given $(J-K_s)$, they are bluer when looking at color
indices that involve $R$ or $I$. We note that only panels (b), (c),
(e) and (f) show colors that include the $H$ band, which hosts the
1.53\,$\mu$m feature itself. The separation between stars showing the 
1.53\,$\mu$m feature and classical C stars, however, is present in all the panels.

This separation into two groups reflects the results of
visual inspection in Sect.\,\ref{section_sample}. Stars with the 1.53\,$\mu$m
feature tend to have excess flux at the red end of the optical
spectrum, for a given energy distribution at farther NIR wavelengths.

\subsection{Molecular indices versus color}

\begin{figure*}
	\begin{center}
		\includegraphics[trim=30 40 30 90, width=\textwidth]{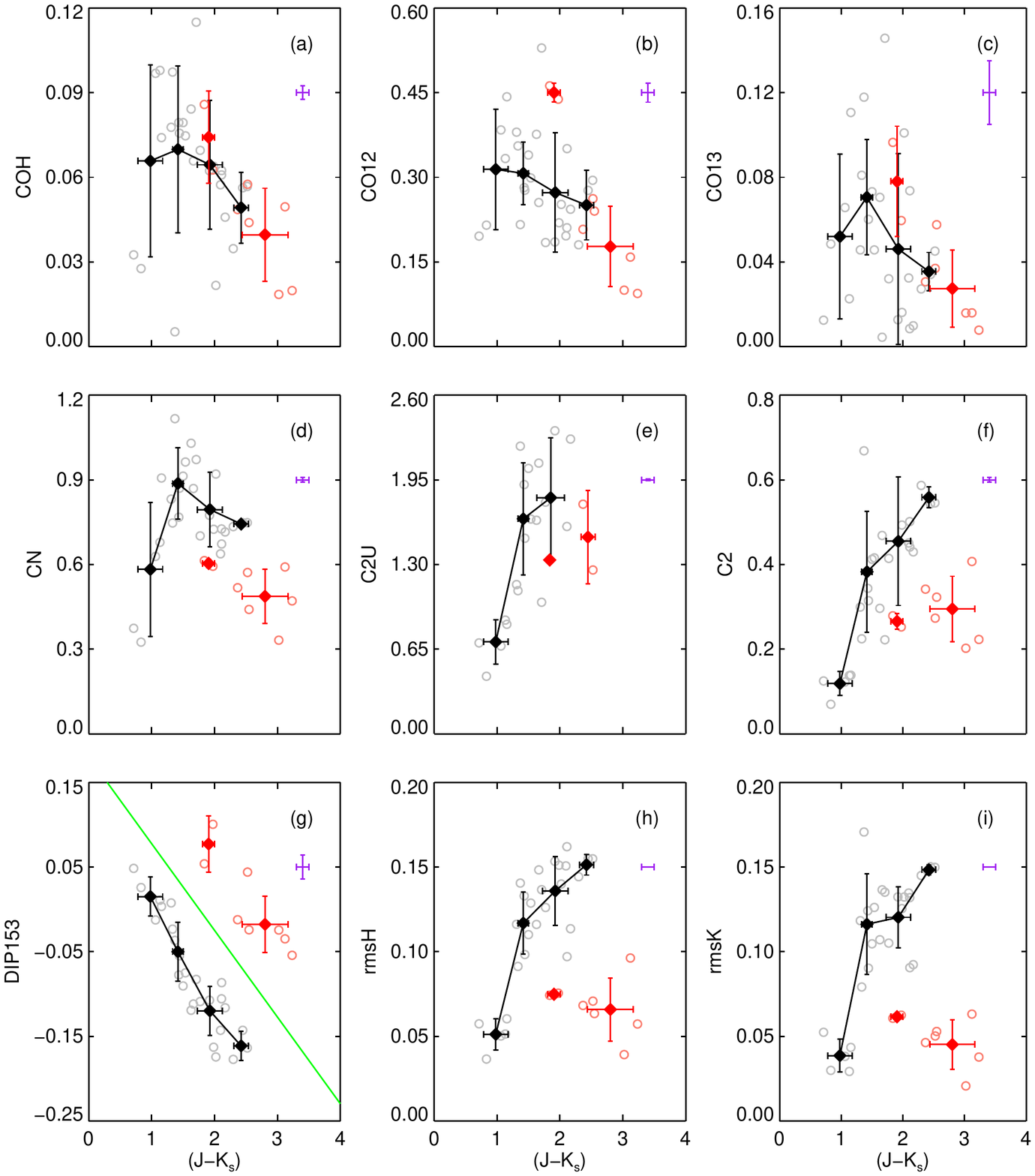}
                \caption{Some indices derived from our sample of
                  carbon stars as a function of $(J-K_s)$. The red
                  circles stand for the carbon stars with the
                  1.53\,$\mu$m feature. The filled diamonds represent
                  the averaged values of our indices as a function of
                  the bin number based on $(J-K_s)$, and the bars measure the dispersion 
                  within bins.
                  Typical uncertainties on individual measurements are shown in purple.}
    \label{plot_color_index}
    	\end{center}
\end{figure*}

Figure~\ref{plot_color_index} shows all the indices previously defined
as a function of $(J-K_s)$. For each group of $(J-K_s)$ (cf. Section~\ref{section_sample}), 
we calculate the average values of each index and display them as filled symbols. 
The error bars indicate the
$1\sigma$ standard deviations. We separate the red C stars from
Groups 3 and 4 into two sub-groups: those with the 1.53\,$\mu$m
feature and those without.

Panels (a) to (c) in Figure~\ref{plot_color_index} display the
\textit{CO} indices as a function of $(J-K_s)$. The data points are
highly dispersed, with a marginal trend of decreasing $^{12}$CO
indices when $(J-K_s)$ increases.

In the $H$-band, the prominence of the CO bands in carbon star spectra
depends on the strength of the CN and C$_2$ bands. These, in turn,
depend on effective temperature, but also on the C/O ratio
\cite[Fig.\,3 of][]{Loidl01}.  While strong in S and S/C stars
\citep{Rayner09,Lancon_Wood}, the $H$-band CO features become
indistinguishable in the forest of CN and C$_2$ lines at large C/O. In
addition, the CO band strengths anticorrelate with surface gravity
\citep{Origlia93}.  In view of the many parameters that may control
the amplitude of the related dispersion of COH values 
(others could include metallicity, microturbulence, convection), 
our sample is too small to consider the trend with $(J-K_s)$
significant.

At 2.29\,$\mu$m, contamination by features other than CO is reduced
but not negligible. Again, we believe the trend with $(J-K_s)$ is only
marginal. Anticipating later discussions, it is interesting to note
that two of the stars with the 1.53\,$\mu$m feature are among those with the
highest values of CO12.

Our measurements of $^{13}$CO approach 0 above
$(J-K_s)\simeq 2$, and its detection is only significant in a few
relatively blue stars. To some extent, this results from contamination
of the measurement bandpasses by a forest of lines from other
molecules and to measurement uncertainties beyond 2.25\,$\mu$m. But
weak $^{13}$C abundances are also a natural and well known result of
the third dredge-up, which brings freshly synthesized $^{12}$C to the
surface \citep{Evans80,Bessel83}.

Panel (d) shows the result for the \textit{CN} index. The behavior is
not monotonic. First, the strength of the CN index increases, up to
$(J-K_s)$ $\simeq$ 1.6. Then, for redder stars, the CN bandhead slowly
fades. For stars with $(J-K_s)$ $\geq$ 2, the sources with the 1.53\,$\mu$m
absorption feature are clearly separated from the classical carbon stars.

Panels (e) and (f) display the \textit{C$_2$} indices. The
\textit{C2U} index is not defined for the reddest objects in the
sample because of a lack of signal at the relevant short wavelengths.
The strength of both \textit{C$_2$} indices increase with increasing
$(J-K_s)$ and drop down for the reddest $(J-K_s)$ values.
The stars with the 1.53\,$\mu$m absorption feature behave differently 
(weaker C$_2$ bands, for a given $(J-K_s)$).

Panel (g) shows the index measuring the strength of the 1.53\,$\mu$m
feature as a function of $(J-K_s)$. There is a clear se\-pa\-ra\-tion
between stars with the 1.53\,$\mu$m feature (red points) and the
others. We add a green line showing this separation for later
comparisons.

The last two panels (h) and (i) display the results for the measure of
the high-frequency structure in the $H$ and $K$ bands.  The
contribution of observational errors to the measured rms indices is in
general smaller than 0.02.  As expected, the values of the rms
increase with increasing $(J-K_s)$, but stars with the 1.53\, $\mu$m
feature follow the opposite trend. The quantitative assessment supports 
the conclusion drawn from visual inspection (see Figures~\ref{all_carbon_p1} and following).


\section{Comparison with the literature}
\label{section_comparison}

Figures~\ref{plot_comp_color} and~\ref{plot_comp_index} compare
our sample of carbon stars with exis\-ting libraries.


\subsection{Spectra from \citet{Lancon_Wood}}

A widely-used library containing carbon stars is that
constructed by \citet[][hereafter LW2000]{Lancon_Wood}. 
They built a library of spectra of luminous cool stars from 0.5 to 2.5\,$\mu$m, 
which includes 7 carbon stars. In addition, their data include
multiple observations of individual variable stars. One of their
stars, R Lep, is a large-amplitude variable star, which exhibits the
1.53\,$\mu$m feature. The different observations of this star are
represented as filled magenta stars in Figures~\ref{plot_comp_color} and ~\ref{plot_comp_index}.


\subsection{Spectra from \citet{Groenewegen09}}

We also compare our spectra with spectra observed by
\citet{Groenewegen09}. They observed carbon rich AGB stars in the Fornax and Sculptor 
dwarf galaxies. The observations co\-ve\-red the entire $J$- and $H$-band atmospheric windows. This sample
is particularly interesting as their color-selected sample happened
to contain several carbon stars with the 1.53\,$\mu$m feature.
Table~\ref{table_groenewegen09} summarizes properties of their carbon
stars: 2MASS $J$, $H$ and $K_s$ magnitudes and also the strength of the
1.53\,$\mu$m feature, as available from Table 1 of \citet{Groenewegen09}. 
Based on their classification, stars for which
this feature is ``medium,'' ``strong'' or ``extreme'' are represented
as filled blue triangles in Figures~\ref{plot_comp_color} and ~\ref{plot_comp_index}.


\begin{table}
\small
\caption{\label{table_groenewegen09} Carbon stars from \citet{Groenewegen09}.}
\centering
\begin{tabular}{lcccr}
\hline\hline
Star & $J$ & $H$ & $K_s$ & Strength of \\
identifier & 2MASS & 2MASS & 2MASS & the 1.53\,$\mu$m feature \\
\hline

Fornax11 	& 15.034 & 13.981 & 13.261 & absent \\
Fornax13 	& 14.485 & 13.377 & 12.618 & weak \\
Fornax15 	& 15.790 & 14.556 & 13.668 & strong \\
Fornax17 	& 14.745 & 13.689 & 13.072 & weak \\
Fornax20 	& 15.131 & 13.732 & 12.728 & weak \\
Fornax21 	& 15.424 & 14.122 & 13.182 & weak \\
Fornax24 	& 15.601 & 14.162 & 13.167 & weak \\
Fornax25 	& 14.722 & 13.262 & 12.120 & weak\\
Fornax27 	& 14.441 & 13.365 & 12.694 & absent \\
Fornax31 	& 16.052 & 14.483 & 13.315 & medium \\
Fornax32 	& 14.789 & 13.664 & 13.076 & absent \\
Fornax34 	& 16.106 & 14.525 & 12.879 & extreme \\
Fornax-S99  & 14.677 & 13.749 & 13.427 & absent\\
Fornax-S116 & 15.004 & 14.041 & 14.028 & absent \\
Scl6 		& 14.846 & 13.144 & 11.603 & strong \\
Scl-Az1-C 	& 14.713 & 14.040 & 13.871 & weak \\ 
\hline
\end{tabular}
\normalsize
\end{table} 


\subsection{Spectra from IRTF \citep{Rayner09}}

Finally, we compare our carbon stars with those from the IRTF Spectral
Library \citep[][hereafter IRTF]{Rayner09}. 
Their collection counts 8 C-star spectra from 0.8 to 5.0\,$\mu$m, including R~Lep. In Figures~\ref{plot_comp_color} and ~\ref{plot_comp_index}, 
this star is represented as a filled green square.


\subsection{Results}
\label{results_comp}

\begin{figure}
	\begin{center}
		\includegraphics[trim=20 50 90 280, clip, width=\hsize]{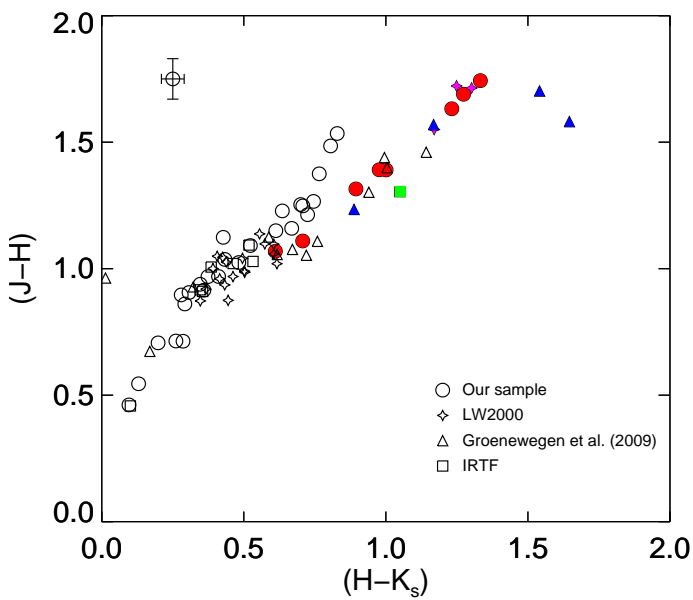}
                \caption{The spread of our sample of carbon stars
                  (circles) in the $(J-H)$--$(H-K_s)$ color-color
                  plane. The star symbols represent LW2000, the
                  triangles represent \citet{Groenewegen09} and the
                  squares represent IRTF. Colored symbols stand for
                  those carbon stars with the 1.53\,$\mu$m feature.
}
    \label{plot_comp_color}
    	\end{center}
\end{figure}

Figure~\ref{plot_comp_color} shows the color--color diagram of $(J-H)$ versus $(H-K_s)$ 
for our sample of carbon stars (circles), to which we add the
C stars from LW2000 (stars), \citet{Groenewegen09} (triangles) and
IRTF (squares). The colored symbols represent the carbon stars with
the 1.53\,$\mu$m feature. All the colors, except those from
\citet{Groenewegen09}, were rescaled to 2MASS values for consistency
in our comparison\footnote{The colors were rescaled by the following factors:\\
$(J-H)_{new}$ = $(J-H)_{ori}$ - 0.10, \\$(H-K_s)_{new}$ = $(H-K_s)_{ori}$ - 0.05, \\ $(J-K_s)_{new}$ = $(J-K_s)_{ori}$ 
- 0.15.}.  
It is interesting to note that the 1.53\,$\mu$m feature makes the 
stars fainter in $H$, so stars with this feature are bluer in $(J-H)$ 
and redder in $(H-K_s)$.

\begin{figure*}
	\begin{center}
		\includegraphics[trim=40 45 0 430]{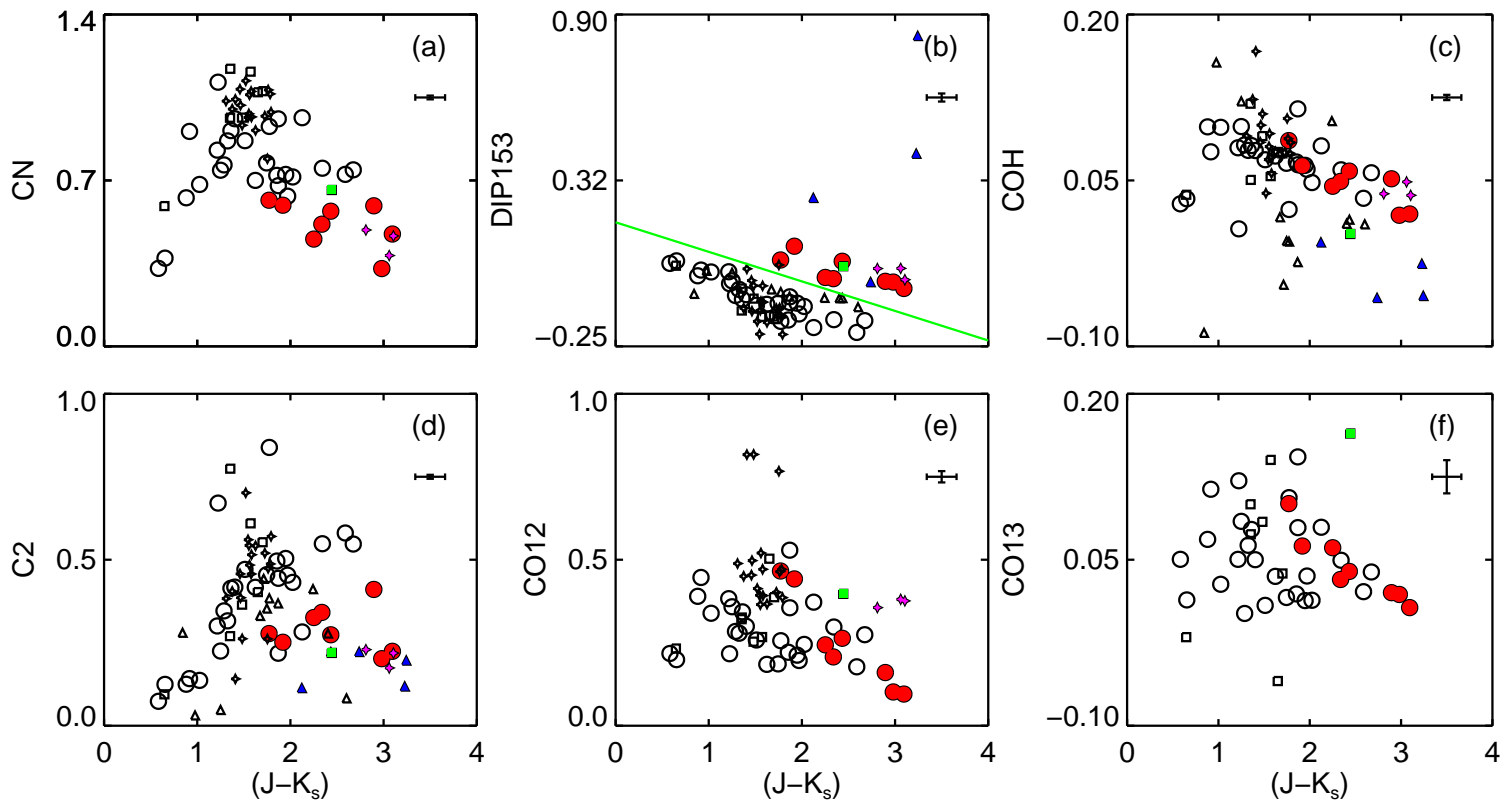}
                \caption{Six of our indices as functions of $(J-K_s)$.
                  Symbols and colors are as in
                  Figure~\ref{plot_comp_color}.  }
    \label{plot_comp_index}
    	\end{center}
\end{figure*}

We compute for the three samples of carbon stars under study the 
spectroscopic indices just like for the XSL spectra. 
Figure~\ref{plot_comp_index} shows six of our indices as a function of $(J-K_s)$. 
Not all the indices are defined for the stars from
\citet{Groenewegen09}, as their spectra are confined to the $J$ and
$H$ bands.

Panel (a) shows the result for the \textit{CN} index. The trend is as
in Figure~\ref{plot_color_index}: an increase of the CN band up to
$(J-K_s)\simeq1.6$ and then a slow decrease at redder colors.

Panel (b) shows the result for the \textit{DIP153} index. The green
line is the same line as in panel (g) of
Figure~\ref{plot_color_index}. Our criterion to identify stars with
the 1.53\,$\mu$m feature appears to be reasonable when applied to a
larger sample.  

In the same panel, one star identified with a star symbol
and not explicitly identified as containing HCN+C$_2$H$_2$ appears
close to the group of stars with the 1.53\,$\mu$m feature. 
This star is BH Cru, a Galactic star, with spectral type CS.
Its IRAS-LRS spectrum shows that it is not
surrounded by a lot of dust \citep[``a hint of circumstellar emission''
as mentioned by][]{Lambert90}. This star displays exceptionally 
strong CO bands in the $H$-band and these contaminate the $DIP153$ index 
(see Figure B28 of \citet{Lancon_Wood}).

Panel (c) plots the \textit{COH} index. We still observe a
decrease of the strength of the COH as $(J-K_s)$ increases.

Panel (d) shows the \textit{C2} index. The most of the ``normal'' stars seem to
lie on a sequence, while the stars with this 1.53\,$\mu$m feature are all
gathered in the lower right-hand quarter of the plot.

Panels (e) and (f) display the \textit{CO12} and \textit{CO13}
indices. The general trend is a decrease of those two indices as
$(J-K_s)$ increases.


\section{Discussion}
\label{section_discussion}

%
\subsection{The bluest stars}

In Section~\ref{blue_stars}, we described the 5 
bluest carbon star spectra
of our sample (Fig.~\ref{all_carbon_p1}).
For the top two stars, HD\,202851 and HE\,1428-195, 
effective temperatures available in the literature are above 4500\,K
\citep{Bergeat02,Placco11}. Hence these are unlikely
to be classical C stars on the AGB. Formation scenarios with mass
transfer from binary companion are more plausible.

The other three stars of the group (Cl*~NGC\,121\,T\,V\,8, 
SHV\,0518161-683543 and SHV\,0517337-725738) cannot be mo\-de\-led 
as reddened versions of the previous two with standard reddening laws.
The energy distributions do not match, and in addition 
circumstellar dust is not expected to play a major role around stars
this warm. These three stars are intrinsically redder than the first two 
and most likely do belong to the AGB.
Only a detailed comparison with models will provide
estimates of their parameters (Gonneau et al., in preparation).
A redder SED can indicate a cooler effective temperature.
Differences in the depth of the molecular bands can be the result of
different metallicities and/or C/N/O abundance ratios. 
Luminosity may also play a role as the strengths of the CO and CN
bands tend to increase when surface gravities drop.

One might be tempted to consider environmental effects, as the bottom
three spectra of Fig.~\ref{all_carbon_p1}
belong to stars in the Large and Small Magellanic Clouds
while the top two are in the Milky Way. 
But one of the two top spectra
belongs to a Galactic Halo star \citep[HE 1428-1950,][]{Goswami10},
and is likely also of subsolar metallicity \citep{Kennedy11}. It is
unknown to which MW population the second top star belongs.

Finally, we note that two of these five spectra clearly display
H$\beta$ in absorption, and two others H$\beta$ in emission. Hydrogen
deficiency is unlikely to be an important characteristic of any of the
stars in this group. Hydrogen emission is a known
transient property of pulsating variables, usually interpreted 
as the results of shocks in the extended atmospheres.


\subsection{The bimodal behavior of red carbon stars}

The previous sections have pointed out a bimodal behavior among
carbon stars with redder near-infrared colors.  The pre\-sen\-ce
or absence of the 1.53\,$\mu$m absorption feature
separates the two types of behaviors (at least in the XSL sample).

\begin{figure}
	\begin{center}
		\includegraphics[trim=65 50 80 70, width=\hsize]{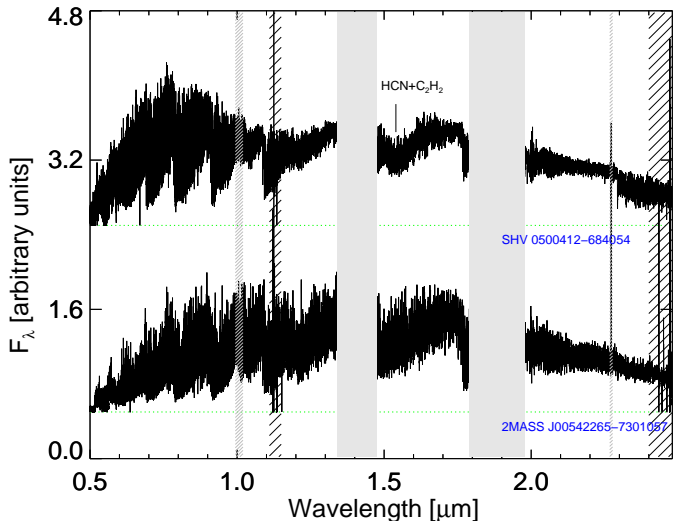}
                \caption{Two high-resolution, high signal-to-noise ratio C-star spectra
                       for which $(J-K_s)\simeq 1.85$.
                       Only the top spectrum exhibits the 1.53\,$\mu$m feature.
                       It also appears smoother in the near-infrared than the bottom one, and it has a peculiar energy distribution.
			The gray bands mask the regions where telluric absorption is strongest. 
                	The areas hatched in black are those that could not be corrected for telluric absorption in a satisfactory way. 
			The a\-reas hatched in gray are the merging regions between the VIS and NIR 
			spectra, and between the last two orders of the NIR part.}
    \label{plot_comp_carbon}
    	\end{center}
\end{figure}

Figure~\ref{plot_comp_carbon} shows two of our high-resolution C-star
spectra with similar $(J-K_s)$ (of $\simeq 1.85$). The 1.53\,$\mu$m
feature is present in the top spectrum but not in the bottom one. In
addition to the HCN+C$_2$H$_2$ feature, the top spectrum shows a
smoother near-infrared spectrum 
and an energy distribution with two components, one peaking at red optical wavelengths, 
the other at long wavelengths.

What might explain this bimodal behavior?  \citet{Groenewegen09} have
already discussed the anti-correlation between the strength of the
1.53\,$\mu$m feature and the C$_2$ bandhead at 1.77\,$\mu$m. As the
1.53\,$\mu$m feature is likely carried in part by C$_2$H$_2$, they
have suggested the formation of this complex molecule occurs at the
expense of C$_2$ when the C/O ratio and the physical conditions are
appropriate. While we have no argument against this suggestion, other
parameters are needed to explain the effects we see in the overall SED
and the anti-correlation with the apparent strength of the forest of
lines in the $H$- and $K$-bands.

Two of the main fundamental parameters of carbon star mo-dels are the
effective temperature and the C/O ratio.  \citet{Aringer09} showed
that dust-free hydrostatic models with any reasonable values
for these parameters fail to reproduce stars redder than $(J-K_s)\simeq 1.6$. 
Circumstellar dust is
invoked to explain redder objects.  At optical and NIR wavelengths,
this dust is usual\-ly thought of as a cause of extinction
\citep{Lancon02,Evans10}.  But extinction by itself cannot explain the
phenomenon we observe.  All classical extinction laws are monotonic
functions of wavelength through the optical and NIR range.  They
cannot explain that, at a given $(J-K_s)$, stars with the 1.53\,$\mu$m
feature have an apparent peak in their $F\!_{\lambda}$
energy distribution in the red part of the optical spectrum.

One explanation for both the SED and the apparent smoothness of
the peculiar spectra in the NIR range (e.g., SHV 0500412-684054) is an
additional continuous {\em emission} component at NIR wavelengths.
While the emission component of the vei\-ling 
by circumstellar dust is usually detected only at longer, 
mid-infrared wavelengths for mass-losing AGB stars, 
the par\-ti\-cu\-lar instantaneous structure around some stars may lead
to a significant contribution in the near-infrared.  
An additive conti\-nuum reduces the equivalent width of absorption 
features in the com\-po\-si\-te spectrum, resulting in a smoother appearance. 
Indeed, such a dilution is seen at NIR wavelengths in some
of the spectra of dusty pulsating C-star models of
\citet{Nowotny11,Nowotny13}, although this aspect was not pointed out
by the authors. Furthermore the existence of objects 
dominated by dust emission at NIR wavelengths such as V CrA (Fig.~\ref{spectrum_vvcra}) 
hints at intermediate situa\-tions.

If we accept the idea of continuous thermal dust emission in the NIR,
we should expect all near-infrared photospheric mo\-le\-cu\-lar 
bands to be similarly weakened. That would explain 
the weaker C$_2$ and CN bands. The stronger HCN+C$_2$H$_2$ absorption 
suggests that these molecules are intermixed with the dust high above the region 
responsible for the other absorption bands. A similar argument was made by \citet{Sloan06} and \citet{Zijlstra06}. 
As the CO bands around 2.3\,$\mu$m do not seem to be weakened 
as much as the C$_2$ or CN lines, CO is also inferred to exist 
at circumstellar radii at least as large as the emitting dust.
This is compatible with dynamical models of carbon star atmospheres, 
which indicate that the robust CO molecule exists 
throughout the circumstellar environment, and the CO(2,0) lines originate
around the radii of dust production, i.e., where most of the 
NIR dust emission is likely to be produced \citep{Nowotny05}.

Indeed, for dust to emit at NIR wavelengths, it must be re\-la\-ti\-ve\-ly
hot and located near the star. This is most likely to happen when the dust first forms, and
to last only until that dust shell is carried farther out into cooler regions. 
Why the presence of dust emission seems to correlate so well with the presence
of the 1.53\,$\mu$m feature would remain to be explained by a renewed
look at models of the chemical structures of pulsating atmospheres.

It is also interesting to note that all the stars in our sample  with 
the 1.53\,$\mu$m absorption band are Miras\footnote{This is also the case 
for the stars with 1.53\,$\mu$m absorption observed by 
\citet{Goebel81}, \citet{Joyce98}, \citet{Lancon_Wood}. 
Among the 9 C stars in \citet{Groenewegen09} with this absorption 
and with some amplitude information \citep[mostly from][]{Whitelock09}
at least four are Miras. Two are labeled
``semi-regular with long term trends'',
three others have quoted $J$-band amplitudes of 0.6, 1.0 and 1.0 in
that article.}
but not all the Miras have this feature (see Table~\ref{table_ext_pp} 
for more details). 
Large-amplitude pulsation might thus be a requirement 
for dust to be forming via a particular chanel, associated with
the NIR signatures of HCN and C$_2$H$_2$ 
and with relatively hot dust temperatures.

\cite{Sloan15} and \cite{Reiter15} recently emphasized 
another example of the connection between the spectral properties
and the pulsation mode of carbon stars: the long period variables 
separate into two sequences when plotting the 
mid-infrared $[5.8]-[8]$ color versus $(J-K_s)$. 
Nearly all semi-regular variables (SRV) follow a blue sequence 
with $(J-K_s) \la 2$, while Miras dominate a redder sequence.
In the range of overlap, around $(J-K_s)\simeq 1.8$, 
SRVs have larger  $[5.8]-[8]$ color indices
than Mira variables, which the authors relate to the presence 
of a strong C$_3$ absorption band around 5\,$\mu$m in SRVs.
The competition between C$_3$, C$_2$, HCN, and C$_2$H$_2$, as well as
the relations between their abundances, their band strengths, the
stellar pulsation amplitude, and the instantaneous atmospheric structure, 
are interesting open questions for future statistical studies.

Finally, we note that this discussion does not 
account for the effects of circumstellar kinematics on line profiles.
Velocity differences larger than 15\,km s$^{-1}$ within the molecular line formation
regions of long-period variables lead to complex broadened line profiles \citep{Nowotny05},
which will affect spectra observed at the resolution of X-shooter.
This has been explored theo\-retically over small wavelength ranges only, and 
at very high spectral resolution. More extensive calculations are
needed to evaluate to what extent the velocity profiles in large-amplitude
variables may contribute to a smoother appearance of some spectra.
However, because the apparent smoothing of spectra with 1.53\,$\mu$m absorption
was detectable in data taken at $R\leq 2000$ \citep{Lancon_Wood,Groenewegen09,Rayner09},
we anticipate that velocity broadening will not by itself explain all the
observations.


\subsection{Synthetic stellar populations with carbon stars}

This collection of spectra of carbon stars offers several improvements
over previous collections for the simulation of spectra of
intermediate age stellar populations. It extends the collection of
\citet{Lancon_Wood} to bluer stars and provides higher resolution at
optical wavelengths. It also offers a new insight into the importance
of evolutionary parameters such as surface composition, pulsation
properties and mass loss along the AGB.

The contributions of TP-AGB and C stars are largest in \mbox{populations} with a
significant component of intermediate-age stars, such as
intermediate-age star clusters
(contributions of the order of 50\% in the $K$ band; \citealt{Ferraro95,Girardi13}) or galaxies with a strong
post-starburst component (\citealt{Lancon99}; \citealt{Miner11}; but
see \citealt{Kriek10} and \citealt{Zibetti13} for counter-examples).
\citet{Melbourne12} use star counts to show the contribution of TP-AGB
stars reaches 17\% at 1.6\,$\mu$m in a sample of local galaxies, but
they do not separate O-rich and C-rich stars.  In near-infrared color-magnitude
diagrams of the Magellanic Clouds, a densely populated plume
of C stars extends from $(J-K)=1.2$ to $(J-K)=2$ at its bright end
\citep{Nikolaev00}.  The correspon\-ding contribution to the light at
2\,$\mu$m amounts to 6 -- 10\% of the total, and is similar to that of
O-rich TP-AGB stars \citep{Melbourne13}.

Many of the C stars detected in the above populations are in the range
of colors for which we have found bimodal spectral properties.  It
will be necessary to evaluate proportions of C stars of the two types
in the future, based on models and observations.  
As we have found that only large-amplitude pulsators in current
samples display the 1.53\,$\mu$m feature and the associated spectral
properties, this implies that future detailed population synthesis
models will need to include the prediction of pulsation properties 
more routinely than is currently the case.
However, consi\-dering the above assessments of the weight of C stars in 
spectra of galaxies, approximate proportions will be sufficient for many purposes.
Demand for precision will come, for instance, when the next generation of
extremely large telescopes will produce detailed near-infrared
observations of star clusters in samples of remote galaxies.

In the short term, the spectra of this paper will be compared with
stellar models, with the double purpose of validating those models and
of constraining the location of the observed stars along stellar
evolution tracks. The X-shooter archive contains additional C-star
spectra that should be added to the XSL collection in the future.
O-rich stars on the AGB are also included in XSL. They should be
compared with those of LW2000 and included in the TP-AGB modeling
procedures.


\section{Conclusions}
\label{section_conclusion}

We have presented a new collection of spectra of carbon stars, 
observed at medium resolution and covering wavelengths from
the near-ultraviolet to the near-infrared.  

Our sample is quite diverse.
We point out a bimodal be\-ha\-vior among the spectra with $(J-K)$ larger than 1.6. 
In addition to the ``classical'' carbon star features, which 
are due pre\-do\-mi\-nantly to CN, C$_2$, and CO,
a subset of our stars also displays an absorption band at
1.53\,$\mu$m. These tend to have a smoother spectral appearance.
In our sample, all the stars with the 1.53\,$\mu$m feature
are large-amplitude (Mira-type) variables. On the other hand, not all the 
Mira-type pulsators display that feature or the associated properties. 

Our interpretation is that 
red carbon stars are enshrouded in circumstellar dust they have produced,
which causes veiling \citep{Nowotny11,Nowotny13}. 
The apparent weakening of the fo\-rest of photospheric molecular lines,
such as those of CN, could occur in specific circumstances, when some of 
the dust is warm enough for the veiling to include a significant 
continuous emission component in the near-infrared. The 1.53\,$\mu$m feature,
which is most likely carried by HCN and C$_2$H$_2$, could be produced in
cool parts of the photospheres or could be distributed within the dusty 
circumstellar environment. 

A comparison with dynamical models, i.e., models in\-clu\-ding pulsation, 
dust formation, and mass loss, is needed to confirm and clarify this picture. Indeed, other mechanisms may
play a role. For instance, the interplay between radiative transfer and
dynamics could matter as it changes individual line profiles
\citep{Nowotny05, Eriksson14}. If the
velocity distributions within the thick photospheres of pulsating
stars are broad enough, a measurable smoothing of the spectra 
could occur. It will also be interesting to compare the near-infrared
fin\-dings with observations at longer wavelengths where  
the spectral properties of carbon stars have also been 
shown to depend on the variability type \citep{Sloan15, Reiter15}.

This collection of spectra will contribute to the improvement 
of stellar population synthesis models. Before using them, it is
necessary to estimate the corresponding
fundamental stellar parameters such as effective 
temperature or surface gravity. 
In a companion paper,
we compare the observations with predictions from hydrostatic models, 
and provide such estimates for spectra which are not affected too 
strongly by pulsation. For strongly pulsating stars, 
high-resolution spectra across a wavelength 
range as broad as that observed with X-shooter remain to be calculated.

\begin{acknowledgements}

  We thank the referee, Dr. Greg Sloan, for the thorough and detailed review that
  helped to improve the quality of this paper. 

  We also thank C. Loup, M. Allen, R. Ibata, M. Groenewegen, and J. van Loon for helpful
  discussions.  We would also like to thank to A. Modigliani, S. Moehler, J. Vinther,
  and the ESO staff for their help during the XSL observations and
  reduction process.  \\
  
  The authors acknowledge financial support from
  PNPS, "Programme National de Physique Stellaire" (Institut National
  des Sciences de l'Univers, CNRS, France) and from NOVA, the
  Netherlands Research School for Astronomy for support. \\
  
  B.~A. acknowledges the support from the {\em project STARKEY} funded by the ERC Consolidator Grant, G.A. n.~615604. \\
  
  This research was funded by the Austrian Science Fund(FWF): P21988-N16. \\
  
  J.~F-B and A.~V acknowledge support from grant AYA2013-48226-C3-1-P from the Spanish Ministry of Economy and Competitiveness (MINECO). \\
  
  This research has made use of the VizieR catalogue access tool and of the
  Simbad database, both operated at CDS, Strasbourg, France. 
  We acknowledge the variable star observations from the AAVSO
  International Database contributed by observers worldwide and used
  in this research, especially the BAAVSS.

\end{acknowledgements}



\bibliographystyle{aa}  
\bibliography{paper_gonneau15_biblio} 



\begin{appendix}


\section{The case of V CrA}
\label{part_vcra}

\begin{figure}
	\begin{center}
		\includegraphics[trim= 70 40 70 70, clip, width=\hsize]{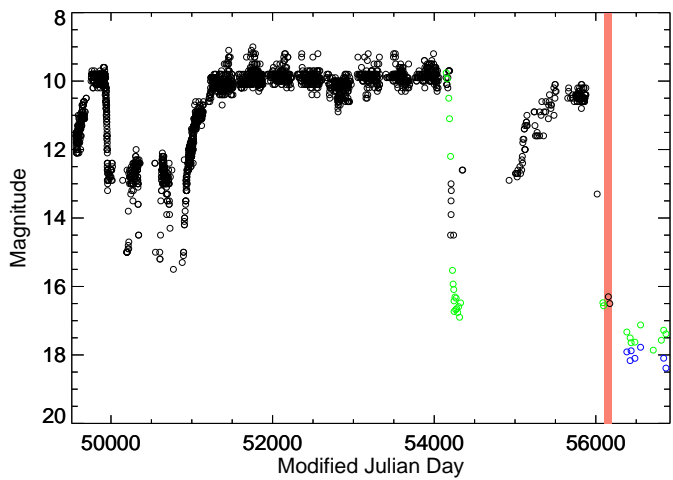}
                \caption{Light curve of V CrA, based on AAVSO data.
                  The time range is from 01/08/1994 to 29/07/2014. Our
                  observing time is highlighted by the red marker and is
        			clearly during a drop in brightness of the star. 
        			The black circles stand for the \textit{Vis} bandpass, 
        			the green points for the $V$ bandpass and 
        			the blue points for the $B$ bandpass.}
	     \label{light_curve_vvcra}
    	\end{center}
\end{figure}

V CrA is a well-known variable star of type R Coronae Borealis (R
CrB).  As discussed by \citet{Evans10}, the characteristic feature of
this star is the occasional decrease in brightness of up to
9 magnitudes, with a rapid drop and a usually slower recovery which
may extend over several years. The common explanation is that the star
ejects clouds of carbon dust which obstruct the light of the star
until they dissipate. Figure~\ref{light_curve_vvcra} shows the light
curve for V CrA over the last 20 years.

\begin{figure}
	\begin{center}
		\includegraphics[trim= 50 50 70 70, clip, width=\hsize]{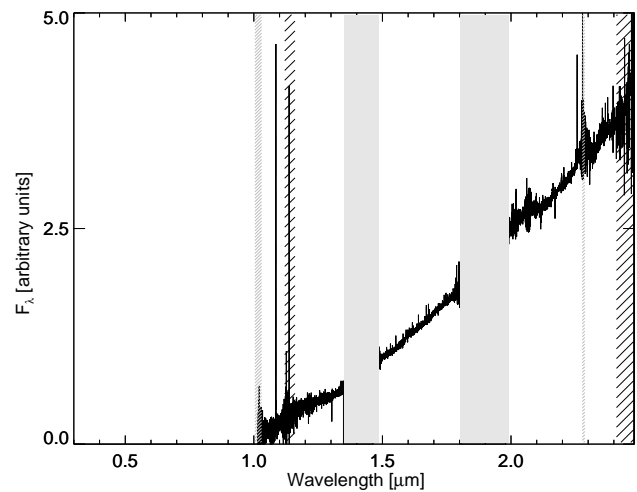}
                \caption{Spectrum of V CrA, from the UVB to NIR. 
                The UVB and VIS parts of the spectrum are consistent with zero flux.
			The gray bands mask the regions where telluric absorption is strongest. 
                	The areas hatched in black are those that could not be corrected for telluric absorption in a satisfactory way. 
					The area hatched in gray is the merging region between the last two orders of the NIR part.}
	     \label{spectrum_vvcra}
    	\end{center}
\end{figure}

We observed V CrA during one of these fading events. Indeed, in its
spectrum, displayed in Figure~\ref{spectrum_vvcra}, we are no longer able
to see the stellar photospheric absorption lines and we only have an emission
spectrum. Because this star is unique in our sample, it is not
considered further in this paper.

\section{Literature properties}

\begin{table*}
\footnotesize
\caption{\label{table_ext_pp} Properties of the observed stars 
{\em (adapted from the literature)}.}

\begin{tabular}{lccccccccccc}
\hline\hline
Name 			& Input  &  Period 	& Ampl. & Band     & Var. &  LC   &  LC  & $K$ 	& $J-K$  & M$_{\mathrm bol}$ & M$_{\mathrm bol}$ \\
     			& cat    &  [d]    	& [mag]	    & \& Ref.  & type &  ref. &  desc. & [mag]	& [mag]  & [mag]  &  ref.   \\
(1)  			& (2)    & (3)     	& (4)       & (5)      & (6)   & (7)  & (8)    & (9)     & (10)  &  (11)  &  (12)  \\ 
\hline
\hline
Cl* NGC 121 T V8        & h      &        	&   	    &          &      &        &        & 12.73 	& 0.93 	&  $-$3.76  	& h    \\
2MASS J00490032-7322238 & g      & 235: 	& 0.40	 & $I$, g, l   & SR   &      l &  irr   & 11.37 	& 1.30 	&  $-$4.70   	&  s \\
2MASS J00493262-7317523 & g      & 175: 	& 0.50	& $I$, g, l    & SR   &      l &  irr   & 10.66 	& 1.39 	&  $-$5.48   	& g \\
2MASS J00530765-7307477 & g      & 948: 	& 0.50 	& $I$, g, l    & SR   &      l &  irr.d & 11.19 	& 1.27 	&  $-$4.89   	& g  \\
2MASS J00542265-7301057 & g      & 1328: 	& 0.80 	& $I$, g, l    & SR   &      l &  irr   & 10.68 	& 1.87 	&  $-$5.19   	& g  \\
2MASS J00553091-7310186 & g      & 216:  	& 1.00 	& $I$, g, l    & M    &      l &  irr.a & 10.56 	& 1.77 	&  $-$5.14   	& g  \\
2MASS J00563906-7304529 & g      & 809: 	& 0.40 	& $I$, g, l    & SR   &      l &  irr   & 11.91 	& 1.24 	&  $-$4.14   	& g  \\
2MASS J00564478-7314347 & g      & 304: 	& 0.60 	& $I$, g, l    & SR   &      l &  irr.a & 10.89 	& 1.47 	&  $-$4.96   	& g  \\
2MASS J00570070-7307505 & g      & 1397: 	& 0.50 	& $I$, g, l    & SR   &      l &  irr   & 10.63 	& 1.65 	&  $-$5.06   	& g  \\
2MASS J00571214-7307045 & g      & 306: 	& 0.40 	& $I$, g, l    & SR   &      l &  irr   & 10.31 	& 1.38 	&  $-$5.51   	& g  \\
2MASS J00571648-7310527 & g      & 913: 	& 0.20 	& $I$, g, l    & SR   &      l &  irr   & 11.38 	& 1.15 	&  $-$4.76   	& g  \\
2MASS J01003150-7307237 & g      & 168: 	& 0.35 	& $I$, g, l    & SR   &      l &  irr.a & 11.86 	& 1.16 	&  $-$4.30        & s  \\
Cl* NGC 419 LE 35       & h      &      	& 0.50 	& $I$, l       &      &        &        & 10.75  	& 1.77 	&  $-$5.02  	& h   \\
Cl* NGC 419 LE 27       & h      &      	& 0.50 	& $I$, l       &      &        &        & 11.00  	& 1.76 	&  $-$5.05  	& h   \\
\hline
T Cae 			& c      & 156: 	& 1.08 	& $V$, i       & SR   & q, o, p & irr   & 2.24 	& 1.52 	&  (5.29)     & s   \\   
\hline
SHV 0500412-684054 	& a 	&  224 	& 0.92 		& $I$, a      &  M  & k, n, m  & reg    & 11.59 	& 1.47 	&  $-$4.53 	& a \\
SHV 0502469-692418 	& a 	&  310 	& 0.92 		& $I$, a      &  M  & k, m     & reg    & 10.65 	& 1.46 	&  $-$4.78 	& a \\
SHV 0504353-712622 	& a 	&  364:	& 1.02		& $I$, a      &  M  & m        & reg.d1 & 10.41 	& 1.96 	&  $-$4.90 	& a \\
SHV 0517337-725738 	& a 	&  152 	& 0.92 		& $I$, a      &   M  &         &        & 12.00 	& 1.05 	&  $-$3.21 	& a \\
SHV 0518222-750327 	& a 	&  338 	& 1.54 		& $I$, a      &  M  &          &        & 10.79 	& 1.48 	&  $-$4.48 	& a \\
SHV 0518161-683543 	& a 	&  193 	& 1.22 		& $I$, a      &  M  & k, m     & reg    & 11.64 	& 1.71 	&  $-$3.75 	& a \\
SHV 0520505-705019 	& a 	&  298 	& 1.20 		& $I$, a      &  M  & k, n, m  & reg.at & 10.99 	& 2.23 	&  $-$4.31 	& a \\
SHV 0520427-693637 	& a 	&  317 	& 1.18 		& $I$, a      &  M  & k, n, m  & reg.atV & 10.77 	& 2.07 	&  $-$3.59 	& a \\
SHV 0528537-695119 	& a 	&  353 	& 0.94 		& $I$, a      &  M  & k        & reg.t  & 10.85 	& 2.83 	&  $-$4.13 	& a \\
SHV 0525478-690944 	& a 	&  398 	& 1.20 		& $I$, a      &  M  & k, m     & reg.t  & 10.17 	& 2.26 	&  $-$5.00 	& a \\
SHV 0527072-701238 	& a 	&  312 	& 1.24 		& $I$, a      &  M  & k, m     & reg.atV & 10.77 	& 2.19 	&  $-$4.67 	& a \\
SHV 0536139-701604 	& a 	&  460 	& 1.34 		& $I$, a      &  M  & k, m     & irr    & 19.73  	& 2.20 	&  $-$4.68	 	& a \\
\hline
{[}ABC89] Pup 42 	& f 	&  	& $\sim$0.14 	& $K$, f      &  SR  &         &       & 6.10 	& 2.24 	&  (9.40)  	& s \\
IRAS 09484-6242 	& f 	&  	& $\sim$0.08 	& $K$, f      &  SR  &         &       & 5.56 	& 2.07 	&  (8.80)  	& s \\
{[}W65] c2      	& f 	&  	& $<0.40$+dips? & $K$, f      &  SR  &         &       & 6.22 	& 1.77 	&  (9.30) 	& s  \\
{[}ABC89] Cir 18 	& f 	&  	& $<0.40$	& $K$, f      &  SR  &         &       & 7.21  	& 2.44  &  (10.50)  	& s \\
HE 1428-1950   		& e 	&    	&    		&              &    &          &      & 9.32  	& 0.67 	&  (11.10) 	& s \\
V CrA 		        & b     &    	& 0.3+dips	& $V$,r        &  R\,CrB   & q, r &   & 7.49 	& 1.20 	&  (10.20)  	& s \\
HD 202851 		& b 	&  	& 	        &    	     &  nv   &   b        &       & 7.02 	& 0.68 	&  (8.80)  	& s \\
\hline
\end{tabular}
\\

{\bf Light curve descriptions in Col. (8)}: irr = irregular;  irr.d = irregular with marked dips in brightness;
irr.a = irregular with marked changes in amplitude; reg = mostly regular; 
reg.d1 = mostly regular with one dip near the end of the available time series; 
reg.at = mostly regular with variations in amplitude and mean magnitude
comparable to half the mean amplitude ; reg.atV = same as reg.at but with larger variations
in the OGLE V light curve; reg.t = mostly regular short period pattern, 
but long term trends in the mean that are comparable or larger 
than the amplitude (possibly associated with a secondary long period).
\\

{\bf References in Tab.\,\ref{table_ext_pp}:} \\
{\bf (a)} \citet{Hughes_Wood}. 
{\bf (b)} \citet{Bergeat02}. Stars found non-variable by these authors
based on Hipparcos satellite photometry are labelled nv in Col.\,(6). \\ 
{\bf (c)} \citet{Lancon_Wood}.
{\bf (e)} \citet{Christlieb01}.
{\bf (f)} \citet{Whitelock06}. \\
{\bf (g)} \citet{Cioni03}  with updates from \citet{Raimondo05}. \\ 
{\bf (h)} \citet{Frogel90} with the distance moduli (m-M)=18.90 for SMC cluster NGC\,419
\citep[][in prep.]{Sloan16} and (m-M)=19.06 for NGC\,121 \citep{Glatt08}. \\
{\bf (i)}  General Catalog of Variable Stars (GCVS) \citep{Samus}.  \\ 
{\bf (k)} OGLE LMC data in $V$ and $I$ \citep[][via Vizier]{Soszynski09}. \\
{\bf (l)} OGLE SMC data in $V$ and $I$ \citep[][via Vizier]{Soszynski11}. \\
{\bf (m)} EROS data in $B$ and $R$ \citep{Kim14}. \\
{\bf (n)} MACHO data in $B$ and $R$ \citep[][2001 version via Vizier]{Alcock96}.\\
{\bf (o)} DIRBE data in $J$ and $K$ \citep[][via Vizier]{Price10}. \\
{\bf (p)} Association Fran\c{c}aise des Observateurs d'Etoiles Variables (AFOEV). \\
{\bf (q)} American Association of Variable Star Observers (AAVSO). \\
{\bf (r)} ASAS data in $V$ \citep[][via Vizier]{Pojmanski02}. \\
{\bf (s)} Our estimate based on $(J-K)$ and \citet{Nowotny13} (see text). When
in parentheses, the values are apparent magnitudes.
\\ ~ \\
{\bf NOTES on individual stars:} \\
{\bf HE 1428-1950} is a high galactic latitude star, which might be metal-poor 
and higher gravity than a typical AGB star \citep{Placco11}. \\ 
{\bf HD 202851} was assigned an approximate bolometric luminosity of -1.9
by \citet{Bergeat02}, which is lower than the expected luminosity
of intrinsic C-stars on the thermally pulsing AGB.\\
{\bf [W65]\,c2}: \citet{Whitelock06} note that \citet{Aaronson88} find $K=$7.21, $(J-K)=$1.95,
while \citet{Aaronson85} had values consistent with 2MASS and themselves. This
star may have obscuration events. 
\end{table*}

\normalsize

To facilitate future usage of the X-shooter spectra, we summarize some of the properties
of the C-stars observed, as found in the literature or derived therefrom.

In Table\,\ref{table_ext_pp}, the stars are sorted by RA with horizontal lines
facilitating the separation between SMC, LMC and Milky Way sources. The names
in Col.\,(1) are understood by standard interpreters such as  those of
CDS (Centre de Donn\'ees astronomiques de Strasbourg).
Col.\,(2) provides the reference of the input catalog used in the
initial construction of the X-shooter sample.
Our systematic searches of the literature for known stellar properties
were made using the generic search through the Vizier catalog database at CDS,
as made available via a search link on the Simbad pages of individual
objects. Information from other sources was added on a best-effort
basis.

Spectral type information was found for only a few of the program stars.
The N star T\,Cae \citep{Alksnis01} is labeled
C6.4 by \cite{Sanford44} and C5II by  \citet{Egret80}.
\citet{Alksnis01} assign type R0 to the R\,CrB star
V\,CrA (in its non-obscured phases) and to HD\,202851,
and type R\footnote{C-R stars, as defined in the older carbon-star classification, are
  usually extrinsic carbon stars \citep{Alknis98}.}
 to IRAS\,09484-6242. In the LMC, the Mira variable SHV0504353-712622 was of type C7.5 when observed by
\citet{vanLoon05}.

Columns (3) to (8) of Tab.\,\ref{table_ext_pp} provide pulsation
information. The periods and amplitudes in columns (3) and (4)
are from $I$ band light curves when found, otherwise as
specified in Col. (5). Periods followed by a colon are highly
uncertain, as the corres\-ponding light curves are very irregular
(for most of these stars, a detailed literature search produces
half a dozen of very different periods depending on passband,
method, and time of observation).
The amplitudes given are representative
of the peak-to-peak variations over long time spans
rather than single-period fits of sine functions
(for V\,CrA, the amplitude given does not include the deep
obscuration events).
The SR and M variability types in Col.\,(6) are based on amplitude (as is common
practice), not on the shape of the light curve.
A comment in Col.\,(8) summarizes our visual inspection of the
available light curves (see Col.\,(7) for references).

The $K$ magnitude and the $(J-K)$ color taken from 2MASS \citep{Cutri03} 
are listed in Col.\,(9) and (10).

The bolometric magnitudes in Col.\,(11) are apparent when in
parentheses, absolute otherwise. For LMC stars, they are
taken from \citet{Hughes_Wood},
who use the bolometric corrections of \citet{Wood83} and a distance modulus of 18.5.
For SMC stars, we list the values of \citet{Cioni03} when avai\-lable. They
are based on energy distributions from the $I$ band to 12\,$\mu$m and we
assumed a distance modulus of 18.90 to the SMC.
We adopted (m-M)=18.90 for SMC cluster NGC\,419
\citep[][in prep.]{Sloan16} and (m-M)=19.06 for NGC\,121 \citep{Glatt08}. For these stars,
we use the apparent bolome\-tric magnitudes of \citet{Frogel90},
which are based on the bolometric corrections of \citet{Frogel80}. 
A more modern compilation of bolometric
corrections BC($K$) as a function of $(J-K)$ is shown in \citet{Nowotny13}.
The spread between authors in that data set amounts to about 0.4\,mag.
The relations of \citet{Wood83} and of \citet{Frogel80} lie near
the ave\-rage of the newer compilation. Representative errors on these
are therefore of the order of 0.2\,mag.

We have not found distances to the Milky Way stars that we consider
reliable, and hence we provide only apparent bolome\-tric magnitudes,
based on $K$, $(J-K)$ and bolometric corrections as above.

\section{Details about the NIR extraction}
\label{part_extraction}

The 1-dimensional (1D) spectra were extracted from 2D images outside of the pipeline, 
with a procedure of our own. The underlying idea was to better control the rejection of bad pixels. 

The standard acquisition procedure for NIR spectra of point
sources is a nodding mode, with observations of the target at two
positions (A and B) along the spectrograph slit.  By default the
pipeline combines the A and B images (flat-fielded and rectified) into
a single 2D image in which the central lines contain the sum of the
stellar spectra of A and B, minus the sum of the skies.  This
accumulates the bad pixels of both the A and B positions into the bad
pixel mask of the combined spectrum, sometimes leaving very few good
pixels to work with.  
Instead, we extracted A from (A$-$B) and B from (B$-$A) and combined them subsequently.

The extraction of a 1D spectrum follows the steps of the optimal 
extraction of \citet{Horne86} quite closely, except that no
parametric fits are involved in the determination of the 2D stellar
illumination profile. For simplicity, we start from rectified, 
wavelength calibrated 2D images of individual orders of the spectrograph. 
The inverse variance weighting scheme neglects the correlation between 
neighbouring pixels that results from the geometrical transformation. 
Although this is not statistically optimal, it performs better than no variance-weighting 
and is sufficient for our purpose.

In the following, we will describe the different components of the input 2D spectra ($\lambda$, $x$) as follows: \begin{enumerate}
\item[1)] The flux spectrum: \textbf{D}
\item[2)] The variance spectrum: \textbf{V}
\item[3)] The quality spectrum: \textbf{M}. \textbf{M} is set to 1 for good pixels and to 0 for
bad pixels.
\end{enumerate}

The first steps of the procedure aim at obtaining a rough first guess for the 1D spectrum.

\begin{enumerate}
\item[a.] \textit{Fit and remove the sky background.} \\
The sky background is estimated on one side of the stellar spectrum on the 2D image. At each $\lambda$, we calculate the median over $x$. We then extend this vector over $x$ to create a 2D vector: \textbf{S}. \\
The new image is therefore: \textbf{(D-S)}. \\
The variance of the new image is: \textbf{(V$_D$ + V$_S$)}. 

\item[b.] \textit{Extract the first guess spectrum.} \\
The first guess spectrum \textbf{f$_1$} is obtained by summing the ima-ge along the spatial dimension $x$. Contrary to \citet{Horne86}, we add the \textbf{M} information, because we find the large values of cosmic ray hits more disturbing than zeros. This choice has no impact on the final result thanks to the smoothing implemented in later steps.
\begin{equation}
f_{1,\lambda} = \sum_{x} (D-S)M
\end{equation}

The variance of the standard spectrum \textbf{V$_1$} is calculated as the sum of the variances along the spatial dimension. 
\end{enumerate}

Once the first guess spectrum is extracted, the ``optimal'' extraction can start. 
This procedure includes three main steps (\textit{c,d,e}), 
and an iteration through all these steps (step \textit{f}).

\begin{enumerate}
\item[c.] \textit{Construct the spatial profile image.} \\
One key component of this procedure is the spatial profile \textbf{P}. 
A first estimate of this profile is built by dividing each line \textit{x} of the sky-subtracted image by the current estimate of the 1D stellar spectrum, \textbf{f}. \\
This initial estimate could be noisy and may contain negative values. Therefore, we improve the profile by smoothing in the wavelength direction, assuming that the spatial profile is a slowly varying function of wavelength. We use a median boxcar of 100$\AA$ width. This erases any outlier pixels but
keeps track of residual curvature or broadening of the spectral image at the end of orders, which sometimes results from an imperfect rectification. \\
Then, we enforce positivity in our profile, by replacing the negative values by 0, and we normalize at each $\lambda$ to make sure that the sum over $x$ is unity.

\item[d.] \textit{Mask the cosmic ray hits.} \\
To mask the cosmic ray hits, we evaluate at each wavelength $\lambda$ the following quantity:
$(D-S-fP)^2 / V$. If this quantity is higher than a given threshold $\sigma_{clip}^2$, the pixel with the largest value is rejected. \citeauthor{Horne86} set this threshold to 25, and we set it to 16. \\
Of course, this pixel rejection could be dangerous. Therefore, we reject up to \textit{only} one pixel per wavelength per iteration, and we update consequently the mask \textbf{M}.  

\item[e.] \textit{Extract the optimal spectrum.} \\
The optimal spectrum is obtained as follows:
\begin{equation}
f_\lambda = \frac{\sum_{x}MP(D-S)/V}{\sum_{x}MP^2/V}
\end{equation}
\begin{equation}
var[f_\lambda] = \frac{\sum_{x}MP}{\sum_{x}MP^2/V}
\end{equation}

\item[f.] \textit{Iterate steps c to e.} \\
To complete the procedure, steps \textit{c} through \textit{e} are repeated 
in order to update the spatial profile image \textbf{P} and the bad-pixel mask \textbf{M}, and
to create new optimal spectra $f_\lambda$ and $var[f_\lambda]$.
This iteration lasts until the set of rejected pixels converges (we allow for a maximum of 3 iterations).

\end{enumerate}

\section{Observed spectra of our sample of carbon stars}

\begin{sidewaysfigure*}
	\begin{center}
		\includegraphics[page=1,trim= 100 50 100 50]{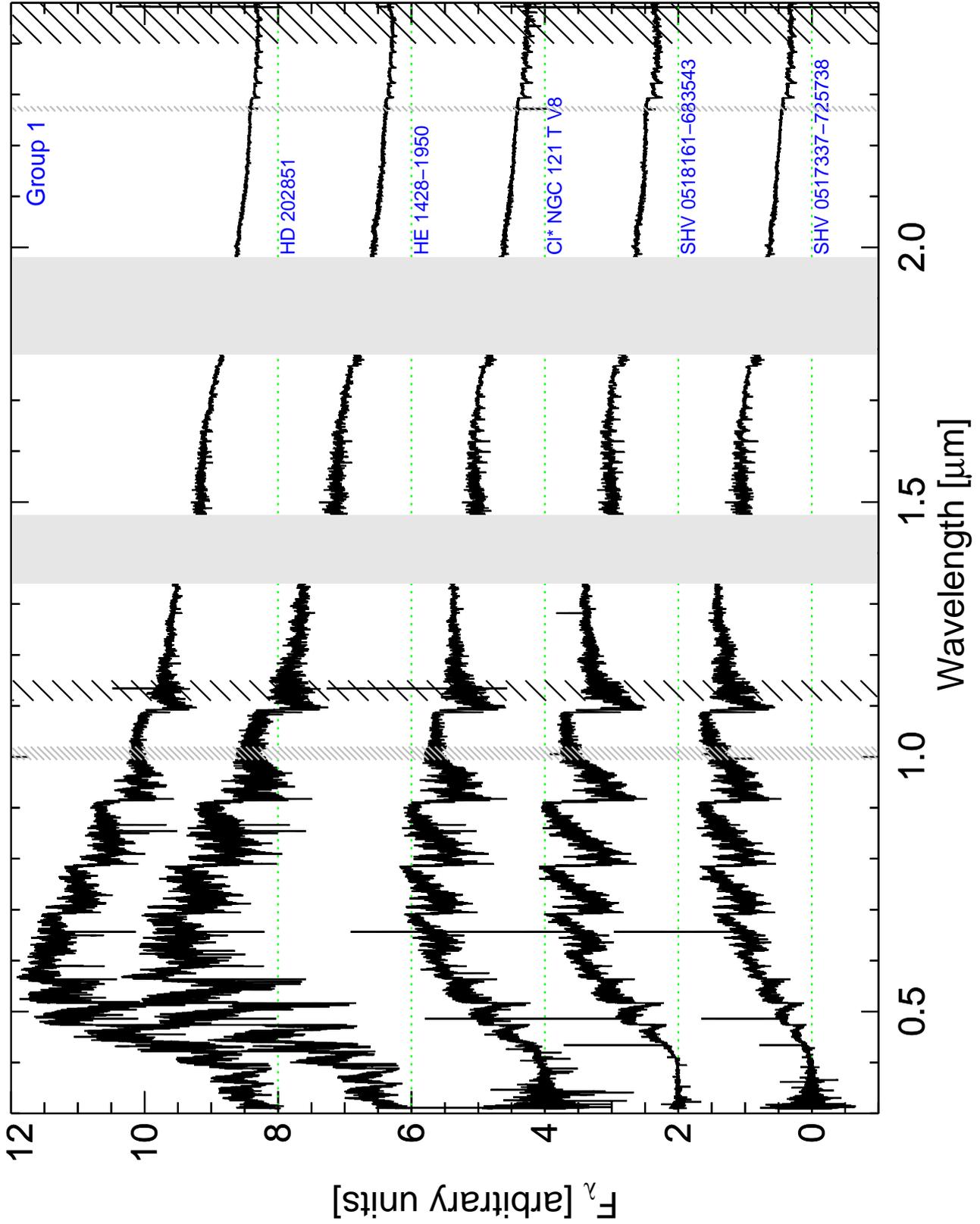}
                \caption{Spectra of our sample of carbon stars. 
					The gray bands mask the regions where telluric absorption is strongest. 
                	The areas hatched in black are those that could not be corrected for telluric absorption in a satisfactory way. 
					The areas hatched in gray are the merging regions between the VIS and NIR spectra and between the last two orders of the NIR part.        
					In some spectra, there is a lack of data at 0.635\,$\mu$m.       	
					The spectra were normalized
                  around 1.7\,$\mu$m and smoothed by the same factor
                  for display purposes (R $\sim$ 2000).}
	     \label{all_carbon_p1}
    	\end{center}
\end{sidewaysfigure*}

Figures~\ref{all_carbon_p1} to~\ref{all_carbon_p6} show the entire
UVB--VIS--NIR spectra for our sample of carbon stars.


\begin{sidewaysfigure*}
	\begin{center}
		\includegraphics[page=2,trim= 100 50 100 50]{spectra_all_carbon}
	     \caption{As for Fig.~\ref{all_carbon_p1}.}
	     \label{all_carbon_p2}
    	\end{center}
\end{sidewaysfigure*}

\begin{sidewaysfigure*}
	\begin{center}
		\includegraphics[page=3,trim= 100 50 100 50]{spectra_all_carbon}
	     \caption{As for Fig.~\ref{all_carbon_p1}.}
	     \label{all_carbon_p3}
    	\end{center}
\end{sidewaysfigure*}

\begin{sidewaysfigure*}
	\begin{center}
		\includegraphics[page=4,trim= 100 50 100 50]{spectra_all_carbon}
	     \caption{As for Fig.~\ref{all_carbon_p1}.}
	     \label{all_carbon_p4}
    	\end{center}
\end{sidewaysfigure*}

\begin{sidewaysfigure*}
	\begin{center}
		\includegraphics[page=5,trim= 100 50 100 50]{spectra_all_carbon}
	     \caption{As for Fig.~\ref{all_carbon_p1}.}
	     \label{all_carbon_p5}
    	\end{center}
\end{sidewaysfigure*}

\begin{sidewaysfigure*}
	\begin{center}
		\includegraphics[page=6,trim= 100 50 100 50]{spectra_all_carbon}
	     \caption{As for Fig.~\ref{all_carbon_p1}.}
	     \label{all_carbon_p6}
    	\end{center}
\end{sidewaysfigure*}


\section{Measurements}

Tables~\ref{table_color_val} and~\ref{table_index_val} display the color and index measurements
associated with Figures~\ref{plot_color_color} and~\ref{plot_color_index}.

\begin{table*}
\caption{\label{table_color_val}Color measurement (as calculated in Section~\ref{section_color}).}
\centering
\small
\begin{tabular}{lccccccc}
\hline\hline

Name & $(R-I)$  & $(R-H)$ & $(I-H)$ & $(I-K_s)$ & $(J-H)$ & $(H-K_s)$ & $(J-K_s)$\\
	& [mag] & [mag] & [mag] & [mag] & [mag] & [mag] & [mag] \\

\hline
\hline

HE 1428-1950 &       0.56 &       1.90 &       1.34 &       1.49 &       0.56 &
      0.15 &       0.71 \\
HD 202851 &       0.50 &       1.95 &       1.45 &       1.63 &       0.65 &
      0.18 &       0.83 \\
Cl* NGC 121 T V8 &       0.74 &       2.65 &       1.91 &       2.16 &
      0.81 &       0.25 &       1.06 \\
SHV 0517337-725738 &       0.91 &       3.16 &       2.26 &       2.57 &
      0.82 &       0.31 &       1.13 \\
SHV 0518161-683543 &       0.75 &       2.78 &       2.03 &       2.37 &
      0.82 &       0.34 &       1.16 \\
2MASS J00571648-7310527 &       0.87 &       3.05 &       2.18 &       2.53 &
      0.96 &       0.35 &       1.31 \\
2MASS J01003150-7307237 &       0.89 &       3.36 &       2.47 &       2.81 &
      1.00 &       0.33 &       1.33 \\
2MASS J00563906-7304529 &       1.34 &       3.98 &       2.63 &       2.99 &
      1.01 &       0.36 &       1.37 \\
2MASS J00530765-7307477 &       0.96 &       3.49 &       2.53 &       2.94 &
      1.02 &       0.41 &       1.43 \\
2MASS J00493262-7317523 &       0.95 &       3.53 &       2.58 &       2.98 &
      1.04 &       0.40 &       1.44 \\
2MASS J00490032-7322238 &       0.93 &       3.39 &       2.45 &       2.88 &
      1.07 &       0.43 &       1.50 \\
2MASS J00571214-7307045 &       0.99 &       3.64 &       2.65 &       3.11 &
      1.07 &       0.46 &       1.54 \\
T Cae &       1.11 &       3.99 &       2.88 &       3.36 &       1.14 &
      0.49 &       1.63 \\
2MASS J00570070-7307505 &       1.14 &       3.74 &       2.60 &       3.14 &
      1.13 &       0.54 &       1.66 \\
{[}W65] c2 &       1.37 &       4.54 &       3.17 &       3.65 &       1.23 &
      0.48 &       1.71 \\
2MASS J00564478-7314347 &       1.15 &       4.37 &       3.22 &       3.79 &
      1.19 &       0.58 &       1.77 \\
SHV 0500412-684054 &       1.01 &       3.52 &       2.51 &       3.17 &
      1.17 &       0.66 &       1.84 \\
2MASS J00542265-7301057 &       1.26 &       4.30 &       3.04 &       3.70 &
      1.25 &       0.67 &       1.92 \\
SHV 0502469-692418 &       1.19 &       4.03 &       2.83 &       3.59 &
      1.21 &       0.76 &       1.97 \\
Cl* NGC 419 LE 27 &       1.72 &       4.67 &       2.95 &       3.67 &
      1.26 &       0.72 &       1.98 \\
IRAS 09484-6242 &       1.38 &       4.46 &       3.08 &       3.76 &       1.33
&       0.69 &       2.02 \\
Cl* NGC 419 LE 35 &       1.71 &       4.70 &       2.99 &       3.77 &
      1.32 &       0.78 &       2.09 \\
2MASS J00553091-7310186 &       1.48 &       5.24 &       3.76 &       4.51 &
      1.36 &       0.75 &       2.11 \\
SHV 0520427-693637 &       1.28 &       4.25 &       2.97 &       3.73 &
      1.35 &       0.76 &       2.11 \\
SHV 0504353-712622 &       1.34 &       4.66 &       3.32 &       4.12 &
      1.37 &       0.80 &       2.17 \\
{[}ABC89] Pup 42 &       1.76 &       5.37 &       3.61 &       4.43 &
      1.48 &       0.82 &       2.30 \\
SHV 0520505-705019 &       1.26 &       4.34 &       3.09 &       4.04 &
      1.42 &       0.95 &       2.37 \\
{[}ABC89] Cir 18 &       1.98 &       6.39 &       4.41 &       5.27 &
      1.59 &       0.86 &       2.45 \\
{[}ABC89] Cir 18 &       1.90 &       6.20 &       4.30 &       5.18 &
      1.64 &       0.88 &       2.52 \\
SHV 0518222-750327 &       1.47 &       5.21 &       3.73 &       4.76 &
      1.49 &       1.03 &       2.52 \\
SHV 0527072-701238 &       1.25 &       4.51 &       3.26 &       4.32 &
      1.49 &       1.05 &       2.55 \\
SHV 0525478-690944 &       2.08 &       7.10 &       5.01 &       6.30 &
      1.73 &       1.29 &       3.02 \\
SHV 0536139-701604 &       1.88 &       6.12 &       4.24 &       5.57 &
      1.79 &       1.33 &       3.12 \\
SHV 0528537-695119 &       0.76 &       4.58 &       3.82 &       5.21 &
      1.85 &       1.39 &       3.23 \\
      
\hline

\end{tabular}
\normalsize
\end{table*}
 
\begin{table*}
\caption{\label{table_index_val}Index measurement (as calculated in Section~\ref{section_indices}).}
\centering
\small
\begin{tabular}{lcccccccccc}
\hline\hline

Name & $(J-K_s)$ [mag] 		& $COH$ & $CO12$ & $CO13$ & $CN$ & $C2U$ & $C2$ & $DIP153$ & $rmsH$ & $rmsK$\\

\hline
\hline

HE 1428-1950 &       0.71 &       0.03 &       0.20 &       0.01 &       0.37 &
      0.70 &       0.13 &       0.05 &       0.06 &       0.05 \\
HD 202851 &       0.83 &       0.03 &       0.22 &       0.05 &       0.33 &
      0.44 &       0.07 &       0.03 &       0.04 &       0.03 \\
Cl* NGC 121 T V8 &       1.06 &       0.10 &       0.38 &       0.07 &
      0.63 &       0.67 &       0.12 &      $-$0.01 &       0.05 &       0.04 \\
SHV 0517337-725738 &       1.13 &       0.10 &       0.33 &       0.02 &
      0.68 &       0.87 &       0.14 &       0.01 &       0.05 &       0.03 \\
SHV 0518161-683543 &       1.16 &       0.07 &       0.44 &       0.11 &
      0.91 &       0.84 &       0.14 &       0.00 &       0.06 &       0.04 \\
2MASS J00571648-7310527 &       1.31 &       0.08 &       0.38 &       0.05 &
      0.83 &       1.15 &       0.30 &       0.01 &       0.12 &       0.12 \\
2MASS J01003150-7307237 &       1.33 &       0.10 &       0.36 &       0.08 &
      0.75 &       1.10 &       0.22 &      $-$0.02 &       0.09 &       0.08 \\
2MASS J00563906-7304529 &       1.37 &       0.01 &       0.22 &       0.12 &
      1.12 &       2.21 &       0.67 &      $-$0.04 &       0.14 &       0.17 \\
2MASS J00530765-7307477 &       1.43 &       0.08 &       0.28 &      0.00 &
      0.77 &       1.91 &       0.34 &      $-$0.08 &       0.13 &       0.12 \\
2MASS J00493262-7317523 &       1.44 &       0.08 &       0.28 &       0.06 &
      0.87 &       1.50 &       0.31 &      $-$0.05 &       0.10 &       0.09 \\
2MASS J00490032-7322238 &       1.50 &       0.08 &       0.34 &       0.07 &
      0.91 &       2.04 &       0.41 &      $-$0.09 &       0.11 &       0.10 \\
2MASS J00571214-7307045 &       1.54 &       0.07 &       0.30 &       0.05 &
      0.96 &       1.65 &       0.42 &      $-$0.07 &       0.13 &       0.13 \\
T Cae &       1.63 &       0.08 &       0.38 &       0.07 &       1.03 &
      1.64 &       0.30 &      $-$0.12 &       0.12 &       0.11 \\
2MASS J00570070-7307505 &       1.66 &       0.07 &       0.26 &       0.00 &
      0.87 &       2.08 &       0.47 &      $-$0.11 &       0.15 &       0.14 \\
{[}W65] c2 &       1.71 &       0.11 &       0.53 &       0.15 &       0.97 &
      1.01 &       0.22 &      $-$0.08 &       0.14 &       0.14 \\
2MASS J00564478-7314347 &       1.77 &       0.07 &       0.18 &       0.03 &
      0.70 &       1.78 &       0.41 &      $-$0.11 &       0.13 &       0.11 \\
SHV 0500412-684054 &       1.84 &       0.09 &       0.46 &       0.10 &
      0.61 &       1.34 &       0.28 &       0.05 &       0.07 &       0.06 \\
2MASS J00542265-7301057 &       1.92 &       0.06 &       0.19 &       0.01 &
      0.77 &       2.33 &       0.45 &      $-$0.11 &       0.15 &       0.13 \\
SHV 0502469-692418 &       1.97 &       0.06 &       0.44 &       0.06 &
      0.59 & - &       0.25 &       0.10 &       0.08 &       0.06 \\
Cl* NGC 419 LE 27 &       1.98 &       0.06 &       0.22 &       0.02 &
      0.72 & - &       0.49 &      $-$0.16 &       0.15 &       0.13 \\
IRAS 09484-6242 &       2.02 &       0.02 &       0.25 &       0.10 &       0.92
& - &       0.83 &      $-$0.17 &       0.14 &       0.13 \\
Cl* NGC 419 LE 35 &       2.09 &       0.06 &       0.20 &       0.03 &
      0.64 & - &       0.45 &      $-$0.14 &       0.15 &       0.13 \\
2MASS J00553091-7310186 &       2.11 &       0.06 &       0.21 &       0.01 &
      0.73 &       1.59 &       0.50 &      $-$0.11 &       0.16 &       0.13 \\
SHV 0520427-693637 &       2.11 &       0.06 &       0.35 &       0.07 &
      0.67 & - &       0.44 &      $-$0.09 &       0.10 &       0.09 \\
SHV 0504353-712622 &       2.17 &       0.05 &       0.24 &       0.01 &
      0.72 &       2.26 &       0.43 &      $-$0.12 &       0.11 &       0.09 \\
{[}ABC89] Pup 42 &       2.30 &       0.03 &       0.18 &       0.03 &
      0.73 & - &       0.59 &      $-$0.18 &       0.14 &       0.15 \\
SHV 0520505-705019 &       2.37 &       0.05 &       0.21 &       0.03 &
      0.52 &       1.76 &       0.34 &      $-$0.01 &       0.07 &       0.05 \\
{[}ABC89] Cir 18 &       2.45 &       0.06 &       0.28 &       0.03 &
      0.75 & - &       0.54 &      $-$0.14 &       0.15 &       0.15 \\
{[}ABC89] Cir 18 &       2.52 &       0.06 &       0.29 &       0.05 &
      0.75 & - &       0.55 &      $-$0.16 &       0.15 &       0.15 \\
SHV 0518222-750327 &       2.52 &       0.06 &       0.26 &       0.04 &
      0.57 &       1.26 &       0.27 &       0.04 &       0.07 &       0.05 \\
SHV 0527072-701238 &       2.55 &       0.04 &       0.24 &       0.06 &
      0.44 & - &       0.32 &      $-$0.02 &       0.06 &       0.05 \\
SHV 0525478-690944 &       3.02 &       0.02 &       0.10 &       0.02 &
      0.33 & - &       0.20 &      $-$0.02 &       0.04 &       0.02 \\
SHV 0536139-701604 &       3.12 &       0.05 &       0.16 &       0.02 &
      0.59 & - &       0.41 &      $-$0.03 &       0.10 &       0.06 \\
SHV 0528537-695119 &       3.23 &       0.02 &       0.09 &       0.01 &
      0.47 & - &       0.22 &      $-$0.05 &       0.06 &       0.04 \\

\hline

\end{tabular}
\normalsize
\end{table*}

Tables~\ref{table_index_lw2000}, ~\ref{table_index_groen} and~\ref{table_index_irtf} displays the color and index measurements for the three comparison samples: \cite{Lancon_Wood}, \cite{Groenewegen09} and \cite{Rayner09}.

\clearpage

\begin{table*}
\caption{\label{table_index_lw2000}Color and index measurements for the data from \cite{Lancon_Wood} (as in Section~\ref{results_comp}).}
\centering
\small
\begin{tabular}{lccccccccc}
\hline\hline

Name & $(J-H)$ & $(H-K_s)$ & $(J-K_s)$ & $CN$ & $DIP153$ & $COH$ & $C2$ & $CO12$ & $CO13$ \\
& [mag] & [mag] & [mag] & & & & & & \\

\hline
\hline

TcaeVK.dec95 &       0.87 &       0.35 &       1.22 &       1.03 &      $-$0.05 &
      0.09 &       0.38 &       0.49 &       0.16 \\
TcaeVK.jan96 &       0.92 &       0.37 &       1.29 &       1.00 &      $-$0.12 &
      0.12 &       0.40 &       0.45 &       0.19 \\
BHcruVK.jan96 &       0.87 &       0.44 &       1.32 &       1.04 &       0.02 &
      0.17 &       0.14 &       0.82 &       0.16 \\
RUpupVK.dec95 &       0.94 &       0.43 &       1.37 &       1.08 &      $-$0.02 &
      0.08 &       0.46 &       0.45 &       0.16 \\
TcaeVK.mar96 &       0.96 &       0.41 &       1.37 &       1.02 &      $-$0.13 &
      0.10 &       0.39 &       0.50 &       0.19 \\
BHcruVK.may96 &       1.00 &       0.39 &       1.39 &       0.93 &      $-$0.04 &
      0.11 &       0.26 &       0.82 &       0.17 \\
ScenVK.jan96 &       0.97 &       0.46 &       1.43 &       1.12 &      $-$0.16 &
      0.04 &       0.70 &       0.41 &       0.24 \\
YhyaVK.may96 &       1.05 &       0.41 &       1.46 &       0.97 &      $-$0.21 &
      0.07 &       0.56 &       0.39 &       0.15 \\
YhyaVK.jun95 &       1.03 &       0.44 &       1.46 &       0.98 &      $-$0.12 &
      0.08 &       0.54 &       0.36 &       0.17 \\
RUpupVK.jun95 &       1.04 &       0.43 &       1.47 &       1.06 &      $-$0.09 &
      0.09 &       0.48 &       0.52 &       0.16 \\
RUpupVK.mar96 &       0.99 &       0.50 &       1.49 &       1.08 &      $-$0.04 &
      0.07 &       0.46 &       0.47 &       0.15 \\
YhyaVK.dec95 &       0.99 &       0.50 &       1.49 &       0.97 &      $-$0.10 &
      0.06 &       0.52 &       0.39 &       0.15 \\
YhyaVK.jul96 &       1.04 &       0.50 &       1.54 &       0.91 &      $-$0.17 &
      0.08 &       0.54 &       0.37 &       0.17 \\
YhyaVK.mar96 &       1.02 &       0.62 &       1.64 &       0.97 &      $-$0.12 &
      0.08 &       0.52 &       0.40 &       0.12 \\
BHcruVK.jul96 &       1.06 &       0.61 &       1.66 &       0.79 &       0.03 &
      0.11 &       0.26 &       0.77 &       0.21 \\
RUpupVK.jan96 &       1.10 &       0.57 &       1.67 &       1.08 &      $-$0.12 &
      0.09 &       0.47 &       0.46 &       0.16 \\
RUpupVK.may96 &       1.14 &       0.56 &       1.69 &       1.06 &      $-$0.15 &
      0.07 &       0.49 &       0.47 &       0.14 \\
YhyaVK.jan96 &       1.10 &       0.61 &       1.70 &       0.99 &      $-$0.21 &
      0.08 &       0.57 &       0.39 &       0.13 \\
RlepVK.dec95 &       1.55 &       1.17 &       2.72 &       0.49 &       0.02 &
      0.04 &       0.23 &       0.36 &       0.12 \\
RlepVK.mar96 &       1.72 &       1.25 &       2.97 &       0.38 &       0.02 &
      0.05 &       0.17 &       0.38 &       0.14 \\
RlepVK.jan96 &       1.71 &       1.30 &       3.02 &       0.47 &      $-$0.02 &
      0.04 &       0.22 &       0.38 &       0.12 \\

\hline

\end{tabular}
\normalsize
\end{table*}
 
\begin{table*}
\caption{\label{table_index_groen}Color and index measurements for the data from \cite{Groenewegen09} (as in Section~\ref{results_comp}).}
\centering
\small
\begin{tabular}{lccccccccc}
\hline\hline

Name & $(J-H)$ & $(H-K_s)$ & $(J-K_s)$ & $CN$ & $DIP153$ & $COH$ & $C2$ & $CO12$ & $CO13$ \\
& [mag] & [mag] & [mag] & & & & & & \\

\hline
\hline
Scl-Az1-C &       0.67 &       0.17 &       0.84 &       0.00 &      $-$0.07 &
     $-$0.09 &       0.28 &       0.00 &       0.00 \\
Fornax-S116 &       0.96 &       0.01 &       0.98 &       0.00 &       0.01 &
      0.16 &       0.03 &       0.00 &       0.00 \\
Fornax-S99 &       0.93 &       0.32 &       1.25 &       0.00 &       0.01 &
      0.12 &       0.05 &       0.00 &       0.00 \\
Fornax17 &       1.06 &       0.62 &       1.67 &       0.00 &      $-$0.05 &
      0.02 &       0.33 &       0.00 &       0.00 \\
Fornax32 &       1.12 &       0.59 &       1.71 &       0.00 &      $-$0.16 &
     $-$0.04 &       0.44 &       0.00 &       0.00 \\
Fornax27 &       1.08 &       0.67 &       1.75 &       0.00 &      $-$0.14 &
     $-$0.00 &       0.35 &       0.00 &       0.00 \\
Fornax11 &       1.05 &       0.72 &       1.77 &       0.00 &      $-$0.06 &
     $-$0.01 &       0.38 &       0.00 &       0.00 \\
Fornax13 &       1.11 &       0.76 &       1.87 &       0.00 &      $-$0.08 &
     $-$0.02 &       0.37 &       0.00 &       0.00 \\
Fornax15 &       1.23 &       0.89 &       2.12 &       0.00 &       0.26 &
     $-$0.01 &       0.11 &       0.00 &       0.00 \\
Fornax21 &       1.30 &       0.94 &       2.24 &       0.00 &      $-$0.08 &
      0.10 &       0.41 &       0.00 &       0.00 \\
Fornax20 &       1.40 &       1.00 &       2.40 &       0.00 &      $-$0.08 &
      0.01 &       0.28 &       0.00 &       0.00 \\
Fornax24 &       1.44 &       0.99 &       2.43 &       0.00 &      $-$0.08 &
      0.01 &       0.22 &       0.00 &       0.00 \\
Fornax25 &       1.46 &       1.14 &       2.60 &       0.00 &      $-$0.11 &
      0.01 &       0.08 &       0.00 &       0.00 \\
Fornax31 &       1.57 &       1.17 &       2.74 &       0.00 &      $-$0.03 &
     $-$0.06 &       0.22 &       0.00 &       0.00 \\
Fornax34 &       1.58 &       1.65 &       3.23 &       0.00 &       0.42 &
     $-$0.03 &       0.12 &       0.00 &       0.00 \\
Scl6 &       1.70 &       1.54 &       3.24 &       0.00 &       0.83 &
     $-$0.05 &       0.20 &       0.00 &       0.00 \\

\hline

\end{tabular}
\normalsize
\end{table*}
 
\begin{table*}
\caption{\label{table_index_irtf}Color and index measurements for the data from \cite{Rayner09} (as in Section~\ref{results_comp}).}
\centering
\small
\begin{tabular}{lccccccccc}
\hline\hline

Name & $(J-H)$ & $(H-K_s)$ & $(J-K_s)$ & $CN$ & $DIP153$ & $COH$ & $C2$ & $CO12$ & $CO13$ \\
& [mag] & [mag] & [mag] & & & & & & \\

\hline
\hline

HD 76846 &       0.46 &       0.10 &       0.56 &       0.59 &
      0.03 &       0.04 &       0.09 &       0.23 &      $-$0.02 \\
HD 44984 &       0.92 &       0.35 &       1.26 &       0.96 &
     $-$0.05 &       0.12 &       0.27 &       0.32 &       0.10 \\
HD 70138 &       0.91 &       0.35 &       1.26 &       1.17 &
     $-$0.13 &       0.05 &       0.77 &       0.33 &       0.07 \\
HD 92055 &       1.01 &       0.39 &       1.39 &       0.96 &
     $-$0.08 &       0.09 &       0.36 &       0.25 &       0.08 \\
HD 57160 &       1.02 &       0.46 &       1.48 &       1.16 &
     $-$0.15 &       0.05 &       0.61 &       0.27 &       0.14 \\
HD 76221 &       1.03 &       0.53 &       1.56 &       1.07 &
     $-$0.14 &       0.08 &       0.40 &       0.50 &      $-$0.06 \\
HD 48664 &       1.09 &       0.52 &       1.61 &       1.08 &
     $-$0.15 &       0.07 &       0.55 &       0.39 &       0.04 \\
HD 31996 &       1.30 &       1.05 &       2.35 &       0.66 &
      0.03 &       0.00 &       0.22 &       0.40 &       0.16 \\
      
\hline

\end{tabular}
\normalsize
\end{table*}

\end{appendix}

\end{document}